\documentclass{aa}

\usepackage{graphicx,natbib,amssymb}
\bibpunct{(}{)}{;}{a}{}{,}

\def\Vmicro{$v_{\rm micro}$} 
\def\Teff{$T_{\rm eff}$} 
\def\logg{$\log{g}$}
\def\logZ{[M/H]}
\def\kms{km\,s$^{-1}$}

\def\tauross{$\tau_{\rm Ross}$}

\defcitealias{cvm}{VM}

\begin{document}

\title{New grids of ATLAS9 atmospheres I: Influence of convection treatments 
       on model structure and on observable quantities}
   
\author{U. Heiter\inst{1,2}, F. Kupka\inst{1}, C. van 't Veer-Menneret\inst{3},
	C. Barban\inst{3,4}, W.W. Weiss\inst{1}, M.-J. Goupil\inst{3}, 
	W. Schmidt\inst{1,5}, D. Katz\inst{3} \and R. Garrido\inst{6}
       }

\offprints{C. van\ 't Veer}

\institute{Institut f\"ur Astronomie, Universit\"at Wien,
	   T\"urkenschanzstrasse 17, A-1180 Vienna, Austria;
	   email: lastname@astro.univie.ac.at
           \and
      	   Department of Astronomy, Case Western Reserve University,
      	   10900 Euclid Avenue, Cleveland, OH 44106-7215, USA;
      	   email: ulrike@fafnir.astr.cwru.edu
	   \and
	   Observatoire de Paris-Meudon, 5, Place Jules Janssen, 
	   F-92195 Meudon Cedex, France;
	   email: Claude.VantVeer@obspm.fr, Caroline.Barban@obspm.fr,
	   MarieJo.Goupil@obspm.fr, David.Katz@obspm.fr
	   \and
	   National Solar Observatory, 950 N. Cherry Ave.,
	   Tucson, AZ 85719, USA;
	   email: barban@noao.edu
	   \and
	   Max-Planck-Institut f\"ur Astrophysik, Karl-Schwarzschild-Str. 1,
	   D-85741 Garching, Germany;
	   email: wschmidt@mpa-garching.mpg.de
	   \and
 	   Instituto de Astrofisica de Andalucia, C.S.I.C., Apdo. 3004,
	   18080, Granada, Spain;
	   email: garrido@iaa.es 
}

\date{Received ; accepted }
 
\titlerunning{New grids of model atmospheres I}
 
\authorrunning{U. Heiter et al.}

\abstract{
      We present several new sets of grids of model stellar atmospheres 
      computed with modified versions of the ATLAS9 code. Each individual set 
      consists of several grids of models with different metallicities ranging 
      from \logZ\ $= -$2.0 to +1.0 dex. The grids range from 4000 
      to  10000~K in \Teff\ and from 2.0 to 5.0 dex in \logg.
      The individual sets differ from each other and from previous ones
      essentially in the physics used for the treatment of the 
      convective energy transport, in the higher vertical resolution 
      of the atmospheres and in a finer grid in the (\Teff, \logg) plane.
      These improvements enable the 
      computation of derivatives of color indices accurate enough for
      pulsation mode identification. In addition, we show that the chosen vertical 
      resolution is necessary and sufficient for the purpose of stellar 
      interior modelling. To explain the physical differences
      between the model grids we provide a description of the currently 
      available modifications of ATLAS9 according to their treatment 
      of convection. Our critical analysis of the dependence 
      of the atmospheric structure and observable quantities on convection treatment, 
      vertical resolution and metallicity reveals that spectroscopic and photometric 
      observations are best represented when using an inefficient 
      convection treatment. This conclusion holds whatever convection
      formulation investigated here is used, i.e.\ MLT($\alpha=0.5$), CM and CGM are
      equivalent. We also find that changing the convection
      treatment can lead to a change in the effective temperature estimated from
      Str\"omgren color indices from 200 to 400~K.
}
	
\maketitle

\section{Introduction}

Convective transport of energy in a stellar atmosphere is one of the most 
complex astrophysical problems. Many of the approximations usually 
admitted for the stellar interior, such as diffusive radiative transfer, 
are no longer valid. Moreover, throughout most of a convective stellar 
atmosphere radiative losses are large enough to make convection less efficient
in transporting energy than radiation. Only stars which have a surface convection
zone (CZ) extending deep into the stellar envelope can maintain efficient
convective energy transfer near the bottom of their atmosphere. On the other
hand, inefficient convection appears in all stars near the boundary of
a convection zone close to locally stable regions. The modelling of inefficient 
convection requires a detailed knowledge about the effect of radiative gains and
losses on the fluid flow. The situation is particularly complex for stars which
are cool enough to develop a granulation pattern, such as the sun. In this case,
at identical geometrical depths, vastly different physical conditions may be 
encountered depending on whether upflow in a granule or downflow in an 
intergranular lane is considered. The former may be optically thick while the 
latter is already optically thin, a consequence of the extreme temperature 
sensitivity of the dominant opacity source in the solar photosphere, the 
H$^{-}$ ion \citep[cf.\ also][]{Stein98}.

Currently, only very simple convection models are available for routine
computation of extended grids of model atmospheres, while detailed numerical
simulations are still unaffordable for applications that require the
calculation of many thousands of individual model atmospheres over the HR diagram. 
 
Our intention here is first to review the convection models which are available for 
use together with the popular ATLAS9 model atmosphere code by
\citet{kuruczI,kuruczII} \citep[see also][]{cast97}. We provide an overview on what 
is known about the effects of the different convection treatments on model 
atmosphere structure and consequently on observable quantities.

The second purpose of the present paper is to determine to what extent
the precision of fundamental parameters derived from the observed stellar spectrum,
i.e.\ \Teff, gravity and metallicity depend on the model atmosphere.
	
Another objective is to obtain very accurate colors and more importantly very
accurate derivatives of colors, color indices and limb darkening coefficients. 
These quantities are needed in the procedure of pulsation mode identification which
is the first and a crucial step in any seismological study. Indeed probing the
stellar interior of a pulsating star requires the knowledge of the resonant cavity
within which each mode propagates, i.e.\ the physical nature of the pulsation mode
associated with each observed oscillation frequency. One such procedure is based on
the computation of oscillation amplitude ratios and phase differences which in turn
depend on the variation of the colors with effective temperature and gravity. 
The results of this application of the model atmosphere grids will be presented in
the next papers of this series \citep{Bar2002,Gar2002}.
Finally, due to their enhanced resolution the new model grids are also
useful to improve the outer boundary conditions of stellar structure
calculations \citep{Mont2001,DAntona2002}.

These goals are part of a program performed in the framework of 
preparing the COROT space mission (see \citeauthor{COROT}). 
To achieve these purposes, we have used the ATLAS9 code in several 
versions modified for the convection zone treatment to compute new 
grids of model atmospheres, corresponding fluxes, surface intensities, 
$uvby$ colors, synthetic spectra for some representative lines, and
compared them with relevant observations.
	We have three versions of the ATLAS9 code at our disposal: 
\begin{enumerate}
\item The original version from CDROM13 of Kurucz \citep{kuruczI} in
      which the convection zone is treated using mixing length theory (MLT).
      While ATLAS versions from 5 to 8 remained basically close to the 
      formulations given in \citet{BV58} and in
      \citet{CG68}, some improvements were added in ATLAS9
      \citep[cf.][]{cast97}. In Sect.~\ref{previous} we discuss the reasons for 
      our specific selection among these improvements. 
\item The other two versions were provided by one of the authors (FK) who
      modified the code to include turbulent convection models from
      \citet[ CM]{cm}, and from \citet*[ CGM]{cgm}.
\end{enumerate}
Each convection model has been extensively used in the model atmosphere grid
computations which we describe below. All the convection models are of local
type and thus require the prescription of a characteristic length scale. 
Formally,
it is possible to interchange the different length scales associated with the
convection models. The motivation for doing so and a particular example will
be discussed in the next paper of this series \citep{Kup2002}. 

This paper is organized as follows. In Sect.~\ref{previous} we 
review previous works about the effect of the model structure on theoretical
photometric colors and justify the need for new grids of model atmospheres.
In Sect.~\ref{convection} we describe the specific
different convection treatments used and discuss their physical content.

In Sect.~\ref{grids} we give details of the grid computations. 
In Sect.~\ref{effects} we set out and comment the role of the convection
treatments and convection parameters on the model structure, as well as its
dependence on effective temperature, surface gravity, and metallicity. Finally,
we discuss the consequences on observable quantities such as Balmer line profiles,
flux distributions, and colors.

\section{A need for new grids} \label{previous}

The original grids of model atmospheres and colors based on the ATLAS9 code
were published by \citet{kuruczI}. They were computed using the classical 
mixing length theory. Kurucz chose and fixed the mixing length parameter
$\alpha$, i.e.\ the ratio $l/H_p$ of convective scale length $l$ and local
pressure scale height $H_p$, to be 1.25. He also used a prescription for 
overshooting at the top of the convection zone 
(cf.\ also Sect.~\ref{convection}) 
to achieve a better match between computed and observed solar fluxes
for the range of $\alpha$ considered. The parameters obtained from the
comparison with solar data were used for the entire grids published in
\citet{kuruczI}. These grids have now been superseded by a new set with
a slightly modified prescription of the overshooting treatment
\citep[for details see][]{cast97}. More recently, they have also become available
in electronic form \citep{kuruczII}. 

\citet{cast97} compared Johnson colors and the $(b-y)$ and $c$ indices
from the Str\"omgren system with
colors from grids of model atmospheres based on MLT with and without the
overshooting prescription, and with an identical choice for the mixing length.
Considering different methods of determining \Teff\ they concluded that models
without the overshooting treatment yield more consistent results, while for
the solar case a model with overshooting was favored. 
As a consequence of this study, new grids of models, fluxes
and colors were computed by Castelli without any overshooting for several
metallicities and different microturbulent velocities.
They are available at the Kurucz website (``NOVER'' grids).
\citet{cast99} analysed synthetic Johnson UBV colors from these model
atmosphere grids, all based on MLT with $\alpha$ = 1.25.
She analysed the effect of metallicity and microturbulent velocity
and concluded that the indices are affected by both the convection treatment
and the amount of line blanketing. This has to be considered in parameter
determinations for stars with unknown metallicity.

\citet{kunz97} have used the revised version of model
atmosphere grids of \citet{kuruczII} to provide a new calibration
of Geneva photometry for B to G type stars. Comparing their photometrically
determined \Teff\ and \logg\ with evolutionary tracks for the Hyades they
noticed a systematic trend in \Teff\ below 7000~K and a rather pronounced
``bump'' in \logg\ located in the same region. Both results were considered
to indicate shortcomings in the model atmospheres used for the computations
of the synthetic color indices.

\citet[ SK]{smk} were the first to study the role of different
convection treatments implemented in the ATLAS9 code of \citet{kuruczI,kuruczII}
for the synthetic $uvby$ colors. They compared observed color
indices with synthetic ones computed using two versions of ATLAS9: the original
version of \citet{kuruczI} based on MLT treatment of convection, with
and without the overshooting option, and another version modified
to employ the convection model of \citet{cm,CM92}, known as the
CM model and described in Sect.~\ref{convection}. 
For the MLT they prove that models built with overshooting at the top 
of the convection zone, as illustrated in \citet{cast97}, are
discrepant with the observed color indices. This confirmed similar conclusions
drawn by \citet[ hereafter VM]{cvm} for the case of
Balmer line profiles. SK also showed that the CM models give
results generally superior to those obtained with MLT using $\alpha = 1.25$,
because they are in better overall agreement with the observed indices
$(b-y)_0$ and $c_0$. The metallicity index $m_0$ was found to be the most
discrepant one with observations, the CM models remaining in good agreement
only for stars with \Teff\ larger than 7000~K, but clearly discrepant for solar
type stars. A peculiar feature in the gravity sensitive $c_0$-index for
\Teff\ around 7000~K was found to be present in colors predicted using any
of the convection models investigated, similar to the results found by
\citet{kunz97} for MLT model atmospheres for the Geneva
photometric system. 

A similar investigation to the one of SK for the Str\"omgren photometric system
was done later by \citet{schmw}, but for the Geneva system. Moreover, he extended 
it to the CGM convection model which had meanwhile been implemented into the
ATLAS9 code (see Sect.~\ref{convection}). His main conclusion, similar to the one of SK,
can be summarized as follows: {\it synthetic color indices are more sensitive to
the scale length used than to the particular convection model}. For instance,
a value of $\alpha = 1.25$ yields differences in the colors in comparison with
models where $\alpha = 0.5$ which are much larger than the
difference among CM and CGM models as well as MLT models with
$\alpha = 0.5$. He concluded that a value of $\alpha = 1.25$ does not allow
reproducing the observed photometric colors of late A and F stars. However,
discrepancies were also found for the other convection treatments he had
studied, in agreement with the results of SK on the $uvby$ colors.
 
\citet{heiter} investigated the temperature structure
and observed quantities calculated with different convection models
for two $\lambda$ Bootis stars with (\logZ, \Teff) values of ($-$1, 6800~K)
and ($-$2, 7800~K). They found a smaller difference between
the synthetic colors and fluxes and the observations when using
the CM model or MLT without overshooting compared to MLT with overshooting
($\alpha$ was set to 1.25 for the MLT models). For the cooler one among the two
stars, the inclusion of overshooting changed the C, Ti, Cr, and Fe abundances
derived from high resolution spectra by +0.1~dex. They also compared the UV
fluxes of these stars with IUE and TD1  measurements and found the CM convection
model to yield results in best overall agreement while the discrepancies were
largest for MLT models with $\alpha = 1.25$ with overshooting.

Recently, \citet{Gard99} extended the comparison
of SK to the CGM model for the case of Balmer line profiles. It was found
that differences between model atmospheres based on the CM or CGM convection
treatment, and models based on MLT without overshooting
yield rather similar results, while MLT models with overshooting are
clearly different. A recommendation for a particular model was found to be
possible only for distinct, limited regions in \Teff. Their results indicated
that a more thorough study of the hydrogen line broadening mechanisms is
necessary to draw more reliable conclusions on the convection model, as well
as a larger number of standard stars with more accurately known fundamental
parameters. For cool dwarf stars such as the sun, one source of problems
in matching observed Balmer line profiles with synthetic ones has 
been to neglect the self-broadening (line broadening
due to collisions with neutral hydrogen) in the hydrogen line profile
calculations \citep{Bark2000}. However, this effect is too
weak in A and F stars to explain the extent of the discrepancies found in
matching the Balmer lines H$_{\alpha}$ and H$_{\beta}$ with some of the
model atmospheres for the stars in the above mentioned works.

From these previous works we have thus drawn the following considerations
for our grid computations. First, the overshooting prescription of ATLAS9 
was generally found to be less successful in reproducing observations for A 
to G type stars, even though for solar observations the case is less settled.
Thus, we have decided not to include models computed with this treatment 
in our grids. However, for comparison we computed individual models with 
overshooting (always using the correction by \citet{cast96}) 
for our case studies (Figs.~\ref{Ttau_single}, \ref{BLP_single} and \ref{irradiance}).

Second, it has been found that model atmospheres which predict temperature
gradients closer to the radiative one, i.e.\ where convection is less efficient
than predicted by MLT models with $\alpha > 1$, are in better overall 
agreement with observations. This was first noticed by
\citet{fag93} and, quite independently, for the case of ATLAS9
models by \citetalias{cvm}
where in order to reproduce the sequence of Balmer line profiles of
the sun with the same solar model they had to reduce the value of $\alpha$ of
their MLT model atmospheres down to $0.5$. Similar results were
found by \citet{fag93}, \citetalias{cvm} and \citet{Veer:98} for a large range of metallicities and stars of spectral types between
A5 and G5 where Balmer lines are both strong and primarily sensitive to the
temperature stratification. As shown above, this overall conclusion can also be 
drawn from other
types of measurements such as photometry and is found to hold in particular
for A type stars with \Teff\ larger than 7000~K, while results
for stars with lower \Teff\ were generally more discrepant. Consequently,
we have decided to base the majority of our model grid computations on
convection treatments which predict less efficient convection than the
previous model grids published by \citet{kuruczI,kuruczII}
and \citet{cast99}.

As far as oscillation mode identification procedures are concerned, 
it has been demonstrated that the dependency of the colors on \Teff\ and 
\logg\ is not captured smoothly enough by the standard ATLAS9 models. The
effects of the non smooth behavior of the color and limb darkening coefficient
derivatives are larger than the expected effect used for identifying the modes
\citep{Gar2000}. In order to obtain smooth variations of these quantities, we
have found that it is necessary to compute our model atmospheres with a higher 
resolution in temperature distribution with depth and built finer grids in 
\Teff\ and \logg.

\section{Convection Treatment: MLT and FST versions of ATLAS9}  \label{convection}

\subsection{Mixing length theory (MLT)}

Model atmospheres computed with ATLAS9 are based on the classical assumptions
of stationarity and horizontal homogeneity. With these restrictions only
some of the properties of stellar convection can be taken into account.
ATLAS9 permits to include:  
\begin{enumerate}
\item the thermal contributions of convection to the energy flux through
      the atmosphere;
\item the effect of convective motions on the line opacity due to the
      additional Doppler broadening of spectral lines caused by turbulent
      velocity fluctuations on length scales smaller than the mean free optical
      path. This is achieved by specifying a microturbulent velocity 
      $v_{\rm micro}$ \citep[cf.][]{gray}.
\item Optionally, ATLAS9 permits to account for changes in pressure
      stratification due to a turbulent pressure $p_{\rm turb}$.
\end{enumerate}
The convective energy flux $F_{\rm conv}$ in the different versions of ATLAS
has been computed traditionally with the classical mixing length theory 
\citep[cf.][]{Bier48,Oepik50,BV58,cast97}. Classical MLT includes radiative
cooling of the fluid which is particularly important where convection is most
inefficient, near the boundary of stably stratified layers. 
It requires the specification of a characteristic scale length $l$ which
is prescribed to be a fraction $\alpha$ of the local pressure scale height,
\begin{equation}  \label{Hp_formula}
   H_p = \frac{P}{\rho g} = \frac{l}{\alpha}.
\end{equation} 
$l$ is used to describe the distance which fluid elements can travel before they
dissolve. It also specifies the geometrical size of the fluid elements (``bubbles'')
together with a second parameter, the ratio of the fluid element volume $V$ over its
surface area $A$. The quantity $V/(A l)$ has been changed during upgrades of the
ATLAS code \citep[cf.][]{cast96}. The present choice results in the same convective
efficiency as the original one of \citet{BV58} if slightly smaller values of
$\alpha$ are used, i.e.\ the usual choice of $\alpha = 1.25$
in the grids of \citet{kuruczI, kuruczII} corresponds to an
``$\alpha_{\rm BV}$'' of about 1.4 for A to G type main sequence stars.
A detailed summary of the modifications of MLT as used in the ATLAS code
can be found in \citet{cast96}, together with various numerical coefficients 
which we have kept unaltered.

One strong motivation to apply a more complete description of stellar
turbulent convection stems from the result that low values
of the scale length parameter $\alpha$, e.g.\ 0.5, are required to fit Balmer
line profiles for the sun and other cool dwarfs \citep[VM]{fag93},
while much larger values (between 1 and 2) are necessary to reproduce their 
observed radii \citep{Mor94}.
Likewise, the scale length ratio has to be varied over an even larger domain
($1 < \alpha < 3$) to reproduce the red giant branch in HR diagrams of galactic
open clusters and associations for stars with masses ranging from 1~$M_{\sun}$
to 20~$M_{\sun}$ \citep{SC95,SC97}.
    
\subsection{Full spectrum turbulence (FST) convection models}

An alternative to MLT which can address these problems was
introduced by \citet{cm,CM92} and is referred to as the CM convection model.
An improved version was proposed by \citet{cgm} which is known as the CGM
formulation. A main intention behind both models was to improve the physical
description of convection while keeping computational expenses as low as for
MLT. Both models achieve this goal by providing a gradient (diffusion)
approximation for the convective (enthalpy) flux:
\begin{equation}  \label{Fc_formula}
   F_{\rm conv} = K_{\rm t} \beta = 
   K_{\rm rad} T H_p^{-1} (\nabla-\nabla_{\rm ad})\Phi(S),
\end{equation}
where $K_{\rm rad} = 4 a c T^3 / (3 \kappa \rho)$ is the radiative conductivity,
$\Phi = K_{\rm t} / K_{\rm rad}$ is the ratio of turbulent to radiative
conductivity, and
\begin{equation}
   \beta = -\left(\frac{dT}{dz}-\left(\frac{dT}{dz}\right)_{\rm ad}\right) = 
          T H_p^{-1} \left(\nabla-\nabla_{\rm ad}\right)
\end{equation}
is the superadiabatic gradient. The convective efficiency $S$ is given by
\begin{equation} \label{S_def}
    S = {\rm Ra}\cdot{\rm Pr}
      = \frac{g\alpha_{\rm v}\beta l^4}{\nu\chi}\cdot\frac{\nu}{\chi},
\end{equation}
${\rm Ra}$ and ${\rm Pr}$ are Rayleigh and Prandtl numbers of the convective
flow, $\alpha_{\rm v}$ is the volume expansion coefficient, and the meaning of
the other symbols is standard. We recall here that the thermometric conductivity
$\chi$ is related to the radiative conductivity through $K_{\rm rad} = c_p
\rho \chi$ and that $\nu$ is the kinematic viscosity. The quantity $S$ is a useful
measure of efficiency for flows which feature a very low ${\rm Pr}$ number, as
occurs in stellar convection, and for which hence the detailed dependence on
$\nu$ can be neglected in parameterisations. This is possible because viscous
processes act on much longer timescales than radiation ($t_{\chi} = l^2/\chi$)
and buoyancy ($t_{\rm b} = (g\alpha_{\rm v}\beta)^{-1/2}$) which in turn are
responsible for the energy balance in stellar atmospheres and envelopes. Thus,
the convective efficiency in a star can be characterized using only
$(t_{\chi} / t_{\rm b})^2 = S$. The latter can easily be related to an
efficiency definition more common in astrophysics \citep{CG68} that
uses the quantity
\begin{equation}
  \Gamma = \frac{1}{2}\left((1 + \Sigma)^{1/2}-1\right),
\end{equation}
where
\begin{equation}
   \Sigma = 4 A^2 (\nabla-\nabla_{\rm ad}) = \frac{2}{81} S,\quad
    A = \frac{Q^{1/2} c_p \rho^2 \kappa l^2}{12 a c T^3}
   \sqrt{\frac{g}{2 H_p}},
\end{equation}
and in which 
$Q = T V^{-1}(\partial V/\partial T)_P = 1-(\partial \ln \mu/ \partial \ln T)_P$ 
is the variable average molecular weight. Using
this notation the MLT of \citet{BV58} can be viewed as a phenomenologically
derived prescription to compute $\Phi$ which reads
\begin{eqnarray} \label{PhiMLT}
   \Phi^{\rm MLT} & = & \frac{9}{8} \Sigma^{-1} \left((1+\Sigma)^{1/2}-1\right)^3 \nonumber\\
       & = & \frac{729}{16} S^{-1} \left((1 + \frac{2}{81} S)^{1/2}-1\right)^3,
\end{eqnarray}
as mentioned by \citet{cm} who pointed out that
alternatively the MLT can be understood as a one-eddy approximation made for
the spectrum $E(k)$ of turbulent kinetic energy \citep[see also][]{Lesi:90}.
The latter describes
how the kinetic energy of the velocity field generated by convection is
distributed among different spatial scales $k^{-1}$.
\citet{Can96} has
shown how MLT underestimates the convective flux in the high efficiency regime
($S \gg 1$) while it overestimates $F_{\rm conv}$ in the low efficiency regime
($S \ll 1$).

Both the CM and CGM convection models attempt to overcome the one-eddy approximation 
by using a turbulence model to compute the full spectrum $E(k)$ of 
a turbulent convective flow for a given $S$, but keep the assumption of
horizontal homogeneity and the Boussinesq approximation used in
MLT.
Hence, they are also referred to as {\em full spectrum turbulence (FST)}
convection models.

In the case of the CM convection model, the so-called eddy damped quasi-normal
Markovian (EDQNM) model \citep{Ors77} of turbulence is used to compute $\Phi(S)$.
This model provides a rather detailed treatment of the nonlinear interactions in
a turbulent flow, but requires the specification of a growth rate. The latter was
computed from the linear unstable convective modes. To avoid the solution of the
equations of the turbulence model each time in a stellar code, the results for 
$F_{\rm conv}$ were tabulated in a dimensionless form. This was achieved
by computing the quantity $\Phi({\rm Ra}, {\rm Pr})$ for a large range of
${\rm Ra}$ and ${\rm Pr}$ numbers. For ${\rm Pr} < 10^{-3}$ the function
$\Phi$ was found to saturate. This agrees with the previous remark that 
$S$ is a useful measure of convective efficiency in a star, where ${\rm Pr}$
is even orders of magnitudes lower, and it was hence sufficient to consider
only the results for the lowest ${\rm Pr}$ number for a tabulation of
$\Phi(S)$, or actually $\Phi(\Sigma)$, given by the EDQNM model.
\citet{cm} found that $\Phi(\Sigma)$ can be represented
by the following analytical fit formula to an accuracy of better than 3\%:
\begin{equation}  \label{Phi}
  \Phi^{\rm CM}(\Sigma) = a_1\Sigma^k \left((1 + a_2 \Sigma)^m - 1\right)^n,
  \quad\mbox{where}
\end{equation}
\begin{eqnarray}  \label{coeffCM}
 & & a_1 = 24.868,\, a_2 = 0.097666, \nonumber\\ & & k = 0.14972,\, m = 0.18931,\, n = 1.8503.
\end{eqnarray}
The comparison with the CGM model published later is simplified if one considers
a change of variable from $\Sigma$ to $S$. In that case
\begin{equation}  \label{Phi_CM_S}
  \Phi^{\rm CM}(S) = b_1 S^k \left(( 1 + b_2 S)^m - 1\right)^n,
  \quad\mbox{where}
\end{equation}
\begin{eqnarray}  \label{coeffCM_S}
 & & b_1 = 14.288,\, b_2 = 0.0024115, \nonumber\\ & & \mbox{and $k, m, n$ are the same as above}.
\end{eqnarray}
While the asymptotic behavior of both MLT and CM models are equal, i.e.\ they
fulfill the limiting relations $k+m n \simeq 1/2$ and thus 
\begin{equation} \label{phi11}
  \Phi(S) \sim S^{1/2} ~~~~~~~{\rm for} ~~~~~~~S ~~\gg ~~1
\end{equation}
as well as $k+n \simeq 2$ and hence 
\begin{equation} \label{phi12}
  \Phi(S) \sim S^2 ~~~~~~{\rm for} ~~~~~~ S ~~\ll~~ 1,
\end{equation}
a distinguishing feature of $\Phi^{\rm CM}(S)$ is to yield about 10 times
more flux than (\ref{PhiMLT}) for $S \gg 1$, i.e.\
\begin{equation}  \label{cm_mlt_1}
  \Phi^{\rm CM}(S) \sim 10 \Phi^{\rm MLT}(S) ~~~~~~~{\rm for} ~~~~~~~S ~~\gg ~~1
\end{equation}
while
\begin{equation}  \label{cm_mlt_2}
  \Phi^{\rm CM}(S) \sim 0.1 \Phi^{\rm MLT}(S) ~~~~~~~{\rm for} ~~~~~~~S ~~\ll ~~1.
\end{equation}
The function $\Phi^{\rm CM}$ defined by (\ref{Phi})--(\ref{coeffCM})
(or (\ref{Phi_CM_S})--(\ref{coeffCM_S}))
is only the first ingredient of the ``CM model''. Because $\Phi$ is computed as
a function of local variables (\ref{S_def}), it depends on a characteristic length
scale which cannot be provided by the formalism itself. Following the physical argument
that the Boussinesq approximation leaves no natural unit of length other than the
distance to a boundary and that eddies near the boundary of the convection zone
are smaller than in the middle of the same (stacking),
\citet{cm} proposed to take
\begin{equation}  \label{CM_length}
   l = z
\end{equation}
where $z$ is the distance to the nearest stable layer. The combination of
(\ref{Phi})--(\ref{coeffCM}) and (\ref{CM_length}) has subsequently been called
the ``CM model''. In this form it was implemented by \citet{Kup96} into ATLAS9
and used for the model grid computations presented here, although other
prescriptions of $l$ had been implemented and experimented with as well.

In a subsequent paper, \citet{cgm} proposed a different FST convection model
which avoided the usage of a growth rate. Rather, it was taken into account that
the rate of energy input which feeds the velocity fluctuations and thus keeps
convection from decaying is controlled by both the source of instability (buoyancy)
and by the turbulence it generates.
However, the treatment of the nonlinear interactions had to be more simplified
to keep the analytical model manageable. The equations of the turbulence model
were solved in the limit for low ${\rm Pr}$ numbers. The new self-consistently
computed input rate results in an increase of the convective flux for a given
efficiency $S$ which is largest at intermediate values of $S \sim 300$. For
that reason a more complicated analytical fit formula had to be used to
represent the predictions of the turbulence model to an accuracy better than
3\% for all values of $S$. The CGM expression for $\Phi$ reads
\begin{equation}  \label{PhiCGM}
   \Phi^{\rm CGM} = F_1(S) F_2(S)
\end{equation}
where $F_1(S)$ has the same structural form as in the CM model,
\begin{equation}  \label{Phi_CGM_F1}
   F_1(S) = ({\rm Ko}/1.5)^3 a S^k \left(( 1 + b S)^m - 1\right)^n, 
            \quad \mbox{with}
\end{equation}
\begin{eqnarray}  \label{coeff_CGM_F1}
 & & a = 10.8654,\, b = 0.00489073, \nonumber\\ 
 & & k = 0.149888,\, m = 0.189238,\, n = 1.85011,
\end{eqnarray}
while $F_2(S)$ is given by
\begin{equation}  \label{Phi_CGM_F2}
   F_2(S) = 1 + \frac{c S^p}{1+d S^q} + \frac{e S^r}{1+f S^t},\quad\mbox{with}
\end{equation}
\begin{eqnarray}  \label{coeff_CGM_F2}
 & & c = 0.0108071,\, d = 0.00301208, \nonumber\\
 & & e = 0.000334441,\, f = 0.000125, \nonumber\\  
 & & p = 0.72,\, q = 0.92,\,  r = 1.2,\, t=1.5.
\end{eqnarray}
Here, ${\rm Ko}$ is the Kolmogorov constant which has been taken 1.7 in
all our calculations, a value well inside of the experimental range
\citep{Pras94}. Note that $\Phi^{\rm CGM}$ shows the
same asymptotic behavior as $\Phi^{\rm MLT}$ and $\Phi^{\rm CM}$ in the
limits of $S \ll 1$ and $S \gg 1$. Moreover, the CM and CGM functions $\Phi$
approach these limits in a very similar manner, because $F_2(S)\rightarrow 1$
for both very large and very small $S$, and the power exponents $k,m,n$ of
(\ref{coeffCM}), (\ref{coeffCM_S}), and (\ref{coeff_CGM_F1}) are almost identical.
However, while 
\begin{equation}  \label{cgm_cm}
  \Phi^{\rm CGM}(S) \sim \Phi^{\rm CM}(S) ~~~~~~~{\rm for} ~~~S ~~\gg ~~1,
\end{equation}
the low efficiency results differ, as
\begin{equation}  \label{cgm_mlt}
  \Phi^{\rm CGM}(S) \sim 0.3 \Phi^{\rm MLT}(S) ~~~~~~~{\rm for} ~~~S ~~\ll ~~1
\end{equation}
(cf.\ (\ref{cm_mlt_2})). 
On a logarithmic scale, the low efficiency limit of (\ref{PhiCGM}) 
is almost exactly the average of the fluxes of
(\ref{PhiMLT}) and (\ref{Phi_CM_S})--(\ref{coeffCM_S}). The second difference between
the two FST convection models is the choice of the scale length $l$ which
\citet{cgm} have proposed to be
\begin{equation}  \label{CGM_length}
   l = z + \alpha^* H_{p,{\rm top}}.
\end{equation}
This accounts for the observed fact that convection penetrates into
neighboring stable regions and thus the scale length cannot decay to
zero right at the layer where the stratification becomes stable according
to the Schwarzschild criterion. The additional term in (\ref{CGM_length})
is thus supposed to account for overshooting and provides a possibility for
small adjustments, if exact stellar radii are needed, e.g.\ in helioseismology.
However, the meaning of overshooting in this context must not be confused
with the overshooting option offered by the ATLAS9 code. This point deserves
special attention to which we turn in the following.

\subsection{Length scale parameters and overshooting}

The term $\alpha^* H_{p,{\rm top}}$ in (\ref{CGM_length}) accounts for 
the increase of the efficiency of convection due to convective penetration 
at the boundary between a stably and an unstably stratified region compared to
a rigid boundary, for instance a fixed plate. The stellar scenario thus implies
to increase the scale length $l$ which can no longer be forced to zero as in
(\ref{CM_length}). The total flux within convectively stable layers is still taken
equal to the radiative flux. On the other hand, the overshooting prescription included
in \citet{kuruczI,kuruczII} as illustrated in \citet{cast97} was invented to take into account
that overshooting directly changes the temperature gradient also in a stable region
next to a convection zone. The procedure suggested is to simply smooth out the
convective flux over as much as 0.5~$H_p$ in each direction around the last point
where $\nabla = \nabla_{\rm rad}$. This mimics the well-known property found in 
many numerical simulations \citep[e.g.][]{Hurl86,Hurl94} and in solutions of the
nonlocal Reynolds stress equations \citep{Kup99,KM2002} where $F_{\rm conv} > 0$
even though $\nabla-\nabla_{\rm ad} < 0$ in layers right next to a neighboring
convection zone. A steeply decaying $F_{\rm conv}$ cannot be modeled this way
while the adjacent region where $F_{\rm conv}< 0$ has to be neglected by taking
$\nabla = \nabla_{\rm rad}$. The effect of this flux smoothing procedure of ATLAS9
on the emergent flux is large enough to provide an additional degree of freedom
to improve the match of solar observations by adjusting the smoothing width.

In the CGM model, the parameter $\alpha^*$ of (\ref{CGM_length}) is typically of
order 0.1 and may be slightly changed to compensate for uncertainties in opacities
and in the treatment of convection. Values of 0.08 and 0.09, similar to \citet{cgm},
were used for the different grids presented in Sect.~\ref{grids}. However,
the effect of such small changes is minute. No inconsistencies were found in
a recent work by \citet{Mont2001} when model atmospheres computed with
$\alpha^*$=0.09 were matched on top of stellar envelopes at different
$\tau_{\rm Ross}$, despite a slightly larger value was used in the stellar structure
computations to obtain the correct solar radius when using the most recent opacity
data. On the other hand, using $\alpha^*$ to compensate for the Boussinesq
approximation and various homogeneity assumptions in ATLAS9 by a match to, say,
the entropy jump near the stellar surface as found from numerical simulations
(cf.\ \citealt{Ludwig99} who used a combination of the
CM fluxes (\ref{Phi})--(\ref{coeffCM}) and the scale length (\ref{CGM_length}))
may require larger variations for models very different from the sun. However, such
a procedure cannot bring the temperature gradient of ATLAS9 model atmospheres into
agreement with the simulations. The latter avoid horizontal homogeneity
assumptions but cannot be afforded together with a treatment of 
frequency dependent radiative transfer which is comparably sophisticated 
as that one used in ATLAS9. Hence, emergent fluxes,
spectra, and photometric colors will be different as well. As long as such
a matching procedure is not shown to allow an improved match of fundamental star data over
extended parts of the HR diagram (and thus improving over present models, cf.\ various
publications discussed in Sect.~\ref{previous}), its practical advantages appear
more limited. For that reason, we have preferred to use the CGM model as intended by
its authors and studied grids with a constant $\alpha^*$ which makes them 
suitable to be matched with
stellar structure calculations using the same treatment of convection \citep{Mont2001}.

\subsection{Implementation of FST models into ATLAS9}

In the ATLAS9 implementation of the CGM convection model the quantities
(\ref{PhiCGM})--(\ref{coeff_CGM_F2}) are actually computed as functions of
$\Sigma$. Thus, only minimal changes were necessary in the subroutine
TCORR, which performs the temperature correction, and in CONVEC, which computes
the convective flux, for replacing the CM with the CGM model. TCORR and CONVEC
were also the only subroutines that had to be changed for implementing the CM model
into ATLAS9. The scale length of the CGM model is evaluated in the following way:
\begin{equation}
l = \min (z_{\rm top} + \alpha^* H_{p,{\rm top}}, z_{\rm bottom} + \alpha^* H_{p,{\rm bottom}})
\end{equation}
This choice makes convection slightly more efficient in comparison with
(\ref{CGM_length}) and more consistent with the idea of accounting for
overshooting, as the latter is also expected to occur below  convection zones.
For most model atmospheres we found that the differences between these alternative
prescriptions are either zero or negligibly small, because the temperature gradient
for convection zones which are entirely contained within the atmosphere is
practically radiative while for convection zones extending below the atmosphere
the evaluation of $l$ in a pure model atmosphere code necessarily has to occur
at the top of the convection zone.

We note here that in principle (\ref{Phi})--(\ref{coeffCM})
and (\ref{PhiCGM})--(\ref{coeff_CGM_F2}) could also be used together with the
common scale length $l = \alpha H_p$ with $\alpha < 1$, or other scale lengths.
Results on such calculations will be reported in
\citet{Kup2002}.

\subsection{Turbulent pressure and the optically thin limit}

For the CM model, a prescription for the turbulent pressure was published
as well, although the results were given only for $S \gg 1$ and in tabular
form. In stellar atmospheres, $S \gg 1$ is usually attained only in cool
stars and close to the bottom where the Rosseland mean optical depth
$\tau_{\rm Ross} > 10$. Hence, the ATLAS9 implementation of the CM model
does not account for turbulent pressure.
On the other hand, for the CGM convection model analytical fit formulae
for $v_{\rm turb}$ and $p_{\rm turb}$ were published by \citet{cgm} which
can be used even for $S \ll 1$ and were implemented into ATLAS9 as well.
A number of model atmospheres for A to early M type
dwarfs and for giants were computed with the CGM model with and without
the prescription of $p_{\rm turb}$. Differences were found only for stars with
deep envelope convection zones, although in most cases both $T$ and $P$
changed by less than 0.1\% for $\tau_{\rm Ross} < 5$, and by no more than 0.5\%
to 1\% for $10 < \tau_{\rm Ross} < 100$. As the inclusion of $p_{\rm turb}$
slowed down the convergence of models while spectra and colors remained 
indistinguishable from the case $p_{\rm turb}=0$, all the CM and CGM
model atmospheres grids presented here are computed without a $p_{\rm turb}$,
just as their MLT counterparts. We note that for stellar structure calculations
the change in temperature structure due to $p_{\rm turb}$ may be more important
than for flux predictions derived from ATLAS9 model atmospheres. To avoid
discrepancies with the CGM treatment as used in the model grids a reasonable
compromise is to match model atmospheres and stellar
envelopes at a $\tau_{\rm Ross} \sim 10$.

Following a suggestion by Canuto (private communication) the correction of
\citet{Spieg57} for radiative losses in optically thin media was implemented
for the case of the CM model. However, except for late K and
early M dwarfs, where ATLAS9 models are not reliable any more due to the
dominance of molecular lines, the effects were found to be negligible. The primary
reason for this are the very low values of $F_{\rm conv}$ predicted by the
CM model for $\tau_{\rm Ross} < 2$ for stars with \Teff\ $>4000$~K.
For the CGM model, convection is slightly more efficient, but still the
effects of such a correction are expected to be very small. Therefore, no further
experiments with radiative loss rates were made with FST convection models.
The case is different for MLT where the results are more sensitive to the
different cooling rates of ``optically thin bubbles'', as
\begin{equation}  \label{mlt_cgm_cm}
  \Phi^{\rm MLT}(S) > \Phi^{\rm CGM}(S) > \Phi^{\rm CM}(S)
   ~~{\rm for} ~~S ~\lesssim ~10
\end{equation}
because of (\ref{cm_mlt_2}) and (\ref{cgm_mlt}) and due to the much larger
$l$ of MLT for $\alpha > 1$ if $z < H_p$. A correction of $F_{\rm conv}$ 
for the optically thin
limit is always included in the MLT implementation of ATLAS9 \citep{kuruczI,cast97}.

\section{Model grid computation}  \label{grids}

Two model grids have been computed independently at the Paris and Vienna
observatories. 
 
At the Paris Observatory an automatic procedure was created by
one of the authors (DK).
The procedure is interactive, and allows the computation of grids of 
model atmospheres based on the ATLAS9 code, of Balmer line profiles, surface 
fluxes and intensities, colors and synthetic spectra, all in one run.
The flux and temperature computations are iterated until
the following convergence criteria are satisfied:
the maximum of the flux and flux derivative errors have to be equal to or less than one
and ten percent, respectively. In addition, the maximum of the temperature 
correction has to be equal to or less than one K.

In the MLT case, we started from the original Kurucz grids 
\citep{kuruczI,kuruczII} and recomputed the models by the scaling procedure
of the ATLAS9 code. The thickness of the layers of the
model atmospheres was divided by 2 or 4 in comparison with the original
\citet{kuruczI,kuruczII} models, in order to solve numerical 
instabilities in the iteration procedure for the flux computation, and to
provide more accurate photometric colors (see Sect.~\ref{flux_col} and next paper in
this series).
Models with higher resolution converged faster and smaller flux errors
were achieved.

\begin{table*}
\caption{Atmospheric and computational parameters of the model atmosphere grids.}
\label{grid_tab}
\begin{tabular}{lllll}
\hline\hline
 & \multicolumn{4}{c}{Paris} \\
\hline
           & Min  & Max  & Step & \\
\Teff [K]  & 6000 & 8500 & 250  & \\
\logg      &  2.0 & 4.5  & 0.1  & \\
\hline
\logZ      & \multicolumn{4}{c}{$-$1.0, 0.0, +1.0} \\
& & & & \\
\Vmicro\ [\kms] & \multicolumn{4}{c}{2} \\
\hline     
Convection & MLT        & CGM  & CM & \\
Parameter  & 1.25, 0.5  & 0.08 &    & \\
$\Delta$log~\tauross$^{\rm a}$ & \multicolumn{4}{c}{0.0625 or 0.03125} \\
Number of layers & \multicolumn{4}{c}{143 or 285} \\
\hline\hline
& & & & \\
\multicolumn{5}{l}{$^{\rm a}$ log~\tauross(top) = $-$6.875} \\
\end{tabular}
\hspace{5mm}
\begin{tabular}{lllll}
\hline\hline
\multicolumn{4}{c}{Vienna} \\
\hline
 & Min & Max   & Step \\
 &4000 & 10000 & 200  \\
 & 2.0 & 5.0   & 0.2  \\
\hline
\multicolumn{4}{l}{$-$2.0, $-$1.5, $-$1.0, $-$0.5, $-$0.3, $-$0.2, $-$0.1,} \\
\multicolumn{4}{l}{0.0, +0.1, +0.2, +0.3, +0.5, +1.0} \\
\multicolumn{4}{c}{0$^{\rm b}$, 1$^{\rm b}$, 2, 4} \\
\hline
MLT & CGM & CGM & CM \\
0.5 & \multicolumn{2}{c}{0.09} & \\
0.125 & 0.125 & 0.03125 & 0.03125 \\
 72 & 72  & 288 & 288 \\
\hline\hline
& & & & \\
\multicolumn{5}{l}{$^{\rm b}$ in preparation} \\
\end{tabular}
\end{table*}

The parameters used for these model grids are given in Table~\ref{grid_tab}.
We recall that the metallicity is given in terms of the logarithmic ratio 
between the total number of atoms of each species, except for hydrogen and 
helium, over the number of hydrogen atoms, with respect to the solar metallicity 
defined in the same way. For instance, [M/H]=0.0 and $-$1.0 means that the
opacities entering the model calculations are computed using either solar element
abundances or solar element abundances divided by 10 for all elements other than
hydrogen and helium. The MLT models were computed for two values of $\alpha$,
the original value used by Kurucz $\alpha = 1.25$,
and the lower value $\alpha = 0.5$, chosen for
reasons given in Sects.~\ref{convection} and~\ref{effects}.
  
In the CM and CGM cases, we started from our MLT models with $\alpha = 0.5$,
and computed grids with the same set of parameters. For the CGM 
convection a value of $\alpha^* = 0.08$ was chosen 
(see Sect.~\ref{convection} for a discussion).

At the Vienna Observatory, model grids with several combinations of
convection treatment and vertical resolution were computed for
slightly smaller step sizes in \Teff, larger step sizes in \logg\ and more
\logZ\ values. For MLT models a value of 0.5 has been chosen for $\alpha$.
Convection has been turned off for models
with \Teff$\ge$8600~K, because the convective flux can be neglected for higher
temperatures, as can be seen from Fig.~\ref{Fconv_max}.
As in the Paris grid the uppermost layer is located at log~\tauross = $-$6.875.
The difference of consecutive layers in log~\tauross
is 0.125 and 0.03125 for models with 72 and 288 layers, respectively.
In addition to the model atmospheres, fluxes and colors in 12 systems
have been computed. Furthermore, information on the convergence extracted
from the ATLAS9 output is provided for each model.
The atmospheric and computational parameters are summarized in 
Table~\ref{grid_tab}.

The grid computations were performed with the perl package SMGT (Stellar Model
Grid Tool), described in \citet{schmw}\footnote{A summary is given in the Appendix
and directions for the use of this program can be found at
http://ams.astro.univie.ac.at/\~\ heiter/smgt\_usage\_1.html.}.
This non-interactive program runs ATLAS9
repeatedly until the convergence criteria are satisfied for each model. The
output of ATLAS9 is evaluated directly and selected information is provided for
each model, such as the root mean square (RMS) and maximum values of the flux 
and flux derivative errors, the maximum of the convective to total flux ratio, 
the extension of the
convection zone, and the optical depth where the temperature equals \Teff.  
The grids defined in Table~\ref{grid_tab} are available on CDROM on request
from the authors.

We note here that two different, but overlapping grids of model atmospheres
were computed as there were different applications in mind. The main
motivation for the computation of the Paris grids was to calculate photometric
colors and their derivatives with respect to \Teff\ and \logg, which will be used
in view of seismic applications \citep{Wats:88,Garri90,Balo99}. This required rather
small steps in \logg, but a restricted range for \Teff\ and few metallicity values.
The results of this specific application will be discussed in a subsequent
paper of this series. The Vienna grids, on the other hand, are intended for
general use, which is the reason for choosing intermediate values for the
parameter step sizes and covering as much of the HR diagram as possible.
Examples for already published applications of these grids can be found
in \citet[ see below]{Mont2001} and in \citet{DAntona2002}.

\subsection{Resolution} \label{resolution}

\begin{figure}
\resizebox{\hsize}{!}{\includegraphics[bb=75 70 380 245]{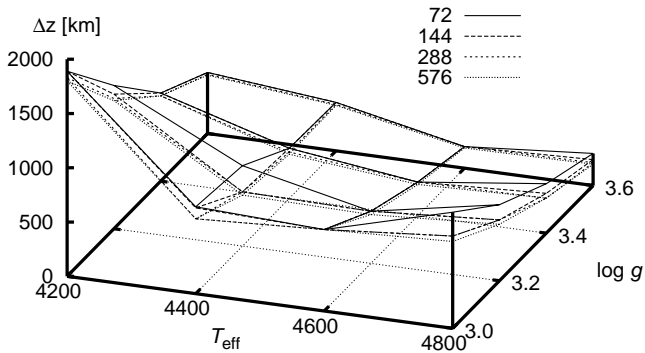}}
  \caption{$\Delta z = z($\tauross$=3.162) - z(F_{\rm conv}=0)$
  for a CGM model grid with four different numbers of layers equally
  distributed between log~\tauross=$-$6.875 and +2. This quantity
  is used for the calculation of the convective scale length in stellar
  interior models and the graph shows its sensitivity to depth resolution
  in this part of the HR diagram.}
  \label{length}
\end{figure}

To show that for specific applications it is necessary to use the models
with 288 layers, we examined the quantity $\Delta z = 
z($\tauross$=3.162) - z(F_{\rm conv}=0)$, where $z$ is the depth
(distance from top layer) in the atmosphere in km and 
$z(F_{\rm conv}=0)$ is the depth of the upper limit of the convection zone. 
This quantity has been used by \citet{Mont2001} for the calculation of
the convective scale length in stellar interior
models which use convective atmospheres computed with ATLAS9 as a
boundary condition. It turned out that for a particular
region in the HR diagram, calculating this
quantity from atmospheric models with 72 layers results in unphysical
oscillations in solar evolutionary tracks which disappear for
higher resolutions (J. Montalb\'an, private communication).
Fig.~\ref{length} shows the values of $\Delta z$ for a small grid
of CGM model atmospheres with [M/H]=0, \Teff=4200\dots4800 and 
\logg=3.0\dots3.6 for four different resolutions,
with the stepsize $\Delta$~log~\tauross\ divided by two for
each successive resolution value. For 144 layers, the results are
rather different from the 72 layer ones (note the peak at (\Teff, \logg)
= (4400, 3.2)). There is a small change when increasing to 288 layers, 
whereas the change when using 576 layers is negligible.
This shows that 288 layers are sufficient in low to moderate temperature
atmospheric models, in particular as all structural quantities
(e.g.\ the temperature gradient $\nabla$) are resolved.
For models with \Teff $\ge$ 10000~K, on the other hand, we verified that
72 layers are sufficient.

\section{The effects of Convection Treatment}  \label{effects}

\subsection{Effects on model atmosphere structure}

\begin{figure*}
\resizebox{\hsize}{!}{\includegraphics[bb=40 150 470 710,clip]{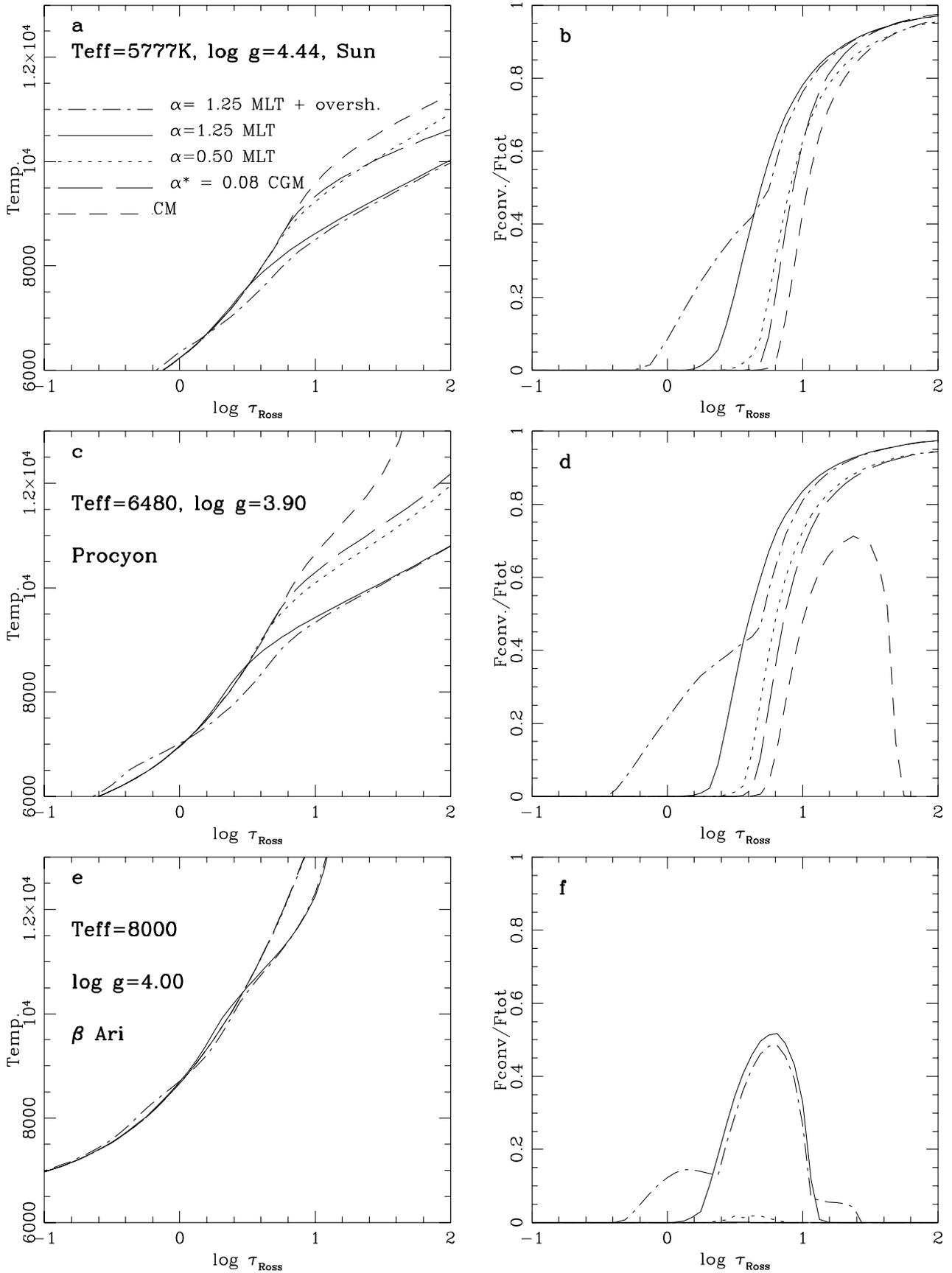}}
  \caption{Distributions of temperature (left panels) and ratio of convective to
  total flux (right panels) with Rosseland optical depth for the three model
  atmospheres adopted for the Sun (a, b), Procyon (c, d) and the A5V star
  $\beta$ Ari (e, f). For each star we show five models computed using different
  convection treatments.}
  \label{Ttau_single}
\end{figure*}

\begin{figure*}
\resizebox{\hsize}{!}{\includegraphics{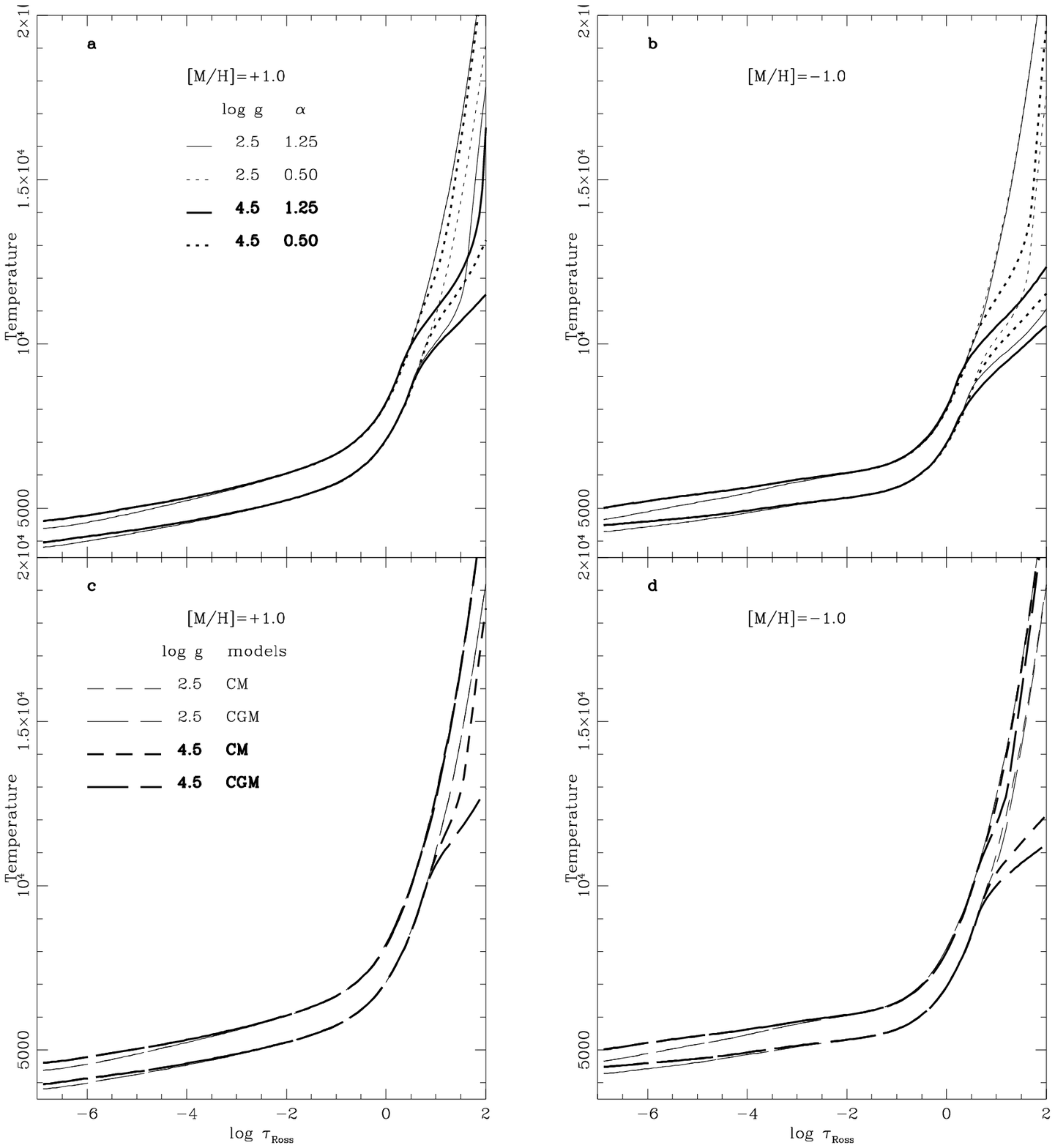}}
  \caption{Temperature versus Rosseland optical depth for models with two different \logg\ values, computed with MLT without overshooting (a: \logZ\ = +1, b: \logZ\ = $-$1, two different values of $\alpha$ for each \logZ) and with FST convection formulation (c: \logZ\ = +1, d: \logZ\ = $-$1, CM and CGM for each \logZ).
  The models are represented for two different values of \Teff\ in each panel, using the same line styles:
  7500~K for the upper four curves and 6500~K for the lower four curves.}
  \label{Ttau_general}
\end{figure*}

\begin{figure}
\resizebox{\hsize}{!}{\includegraphics{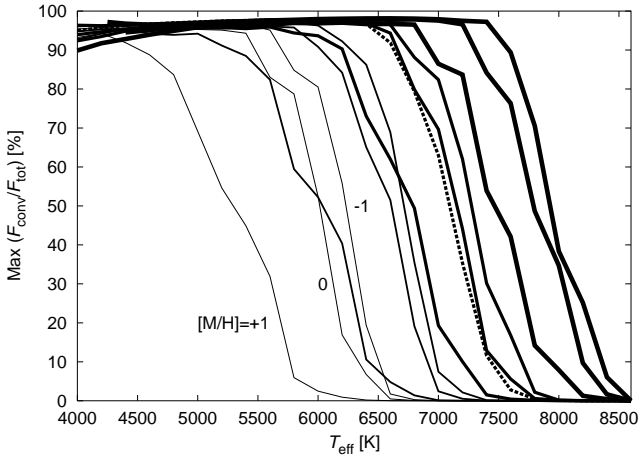}}
  \caption{Maximum of convective to total flux ratios as a function of \Teff, for
  the following values of \logg: 2, 3, 4, 5, represented
  by increasing line thickness, and three different metallicities as labeled for \logg=2,
  with the same trends for the other values. The dashed curve was computed with
  [M/H]=0 and \logg=4, but with a four times higher depth resolution.
  The CGM convection model was used in all cases.}
  \label{Fconv_max}
\end{figure}

We first examine changes of temperature and convective flux distribution
when using different convection models. Figs.~\ref{Ttau_single}, \ref{Ttau_general},
and \ref{Fconv_max} show the intricate dependence of the effect of 
convection treatment on \Teff, \logg, and \logZ\ of the model.

Fig.~\ref{Ttau_single} displays the temperature and the convective flux as
a function of Rosseland optical depth (log~\tauross) corresponding to the models
used for three specific main sequence solar metallicity  stars  which have
been chosen so as to cover the temperature range of interest: the Sun,
Procyon -- a well studied reference star, and $\beta$ Ari -- a well observed
hot star. For each star, several models are computed which differ only for 
the convection treatment. \Teff, \logg, metallicity, and microturbulent velocity
of the models are taken from previous detailed analyses (by CV for $\beta$ Ari, 
\citet{Veer:98} for Procyon and \citet{kuruczII} for the Sun). 

The slope of the $T-\tau$ relation within the CZ indicates the efficiency 
of the convection transport. It is steeper for a less efficient convection, 
i.e.\ a temperature gradient closer to the radiative one. For instance, in Fig.~\ref{Ttau_single}a,
it can be seen that the CM model predicts the least efficient convection, followed
with increasing convective efficiency by the MLT ($\alpha=0.5$), the CGM and 
the MLT ($\alpha=1.25$) models. The same trend is observed for the convective flux
in Fig.~\ref{Ttau_single}b. This is a consequence of the fact that radiative losses of the convective
fluid are always large within the stellar atmosphere where the gas is optically
transparent. Hence, the inequality chain (\ref{mlt_cgm_cm}) always holds
at least at lower optical depths ($ \tau \lesssim 1$). The scale lengths
(\ref{Hp_formula}), (\ref{CGM_length}), (\ref{CM_length}) of MLT, CGM and CM
respectively also obey such an inequality chain for distances $z$ closer to
stably stratified layers than $\alpha H_p$ and for the ranges of $\alpha$ and $\alpha^*$
considered in our work. Therefore, the amount of convective flux and the associated $T-\tau$
relations shown in Fig.~\ref{Ttau_single} are an immediate consequence of these inequality chains.

For hotter stars (\Teff $\simeq 8000 $ K), as the convective efficiency decreases,
all convection models predict a temperature gradient close to the radiative one.

The effect of convection treatment on the atmosphere structure depends on
metallicity, gravity and \Teff\ in a complex way, as illustrated in
Fig.~\ref{Ttau_general}. Evidently, a metal rich atmosphere reduces
the efficiency of the convection transport, as does a low gravity, or a high \Teff.
The influence of \logg\ on the convective efficiency depends strongly
on \Teff, \logZ, and the convection model. For instance, a model at \Teff\ = 6500~K,
\logg\ = 2.5, $\alpha_{\rm MLT}\ = 0.5$, and ten times solar metallicity is
completely radiative (see Fig.~\ref{Ttau_general}a, thin dashed lower line),
while for \logg\ = 4.5 and
identical parameters otherwise a small deviation from radiative stratification
is found (thick dashed lower line). This deviation grows significantly when
decreasing the metallicity to one tenth of the solar one (Fig.~\ref{Ttau_general}b).

The variation of the maxima of the convective flux with \Teff, \logg, and \logZ\ is
shown in Fig.~\ref{Fconv_max} for the CGM models.
The decrease of convective flux with increasing metallicity is a consequence of
the lower mass density ($\rho$) found in metal rich atmospheres.
The latter is a result of the increased opacity, which requires a smaller
column density for a given optical depth.
Due to increased line blanketing in metal rich atmospheres, the requirements of
flux constancy and hydrostatic equilibrium then result in both lower temperature
and lower pressure in the outermost layers.
As a consequence, lower densities are also found near the boundary of the CZ.
This makes convection less efficient, although this effect is partially compensated 
by a higher convective velocity found for metal rich atmospheres.
Fig.~\ref{Fconv_max} also shows the influence of using a four times higher resolution
in optical depth, resulting in a much smoother run of the curves and in 
a small shift towards lower temperatures.

\subsection{Consequences on observable quantities}

\subsubsection{Balmer line profiles}

\begin{figure*}
\resizebox{\hsize}{!}{\includegraphics[bb=50 150 490 710]{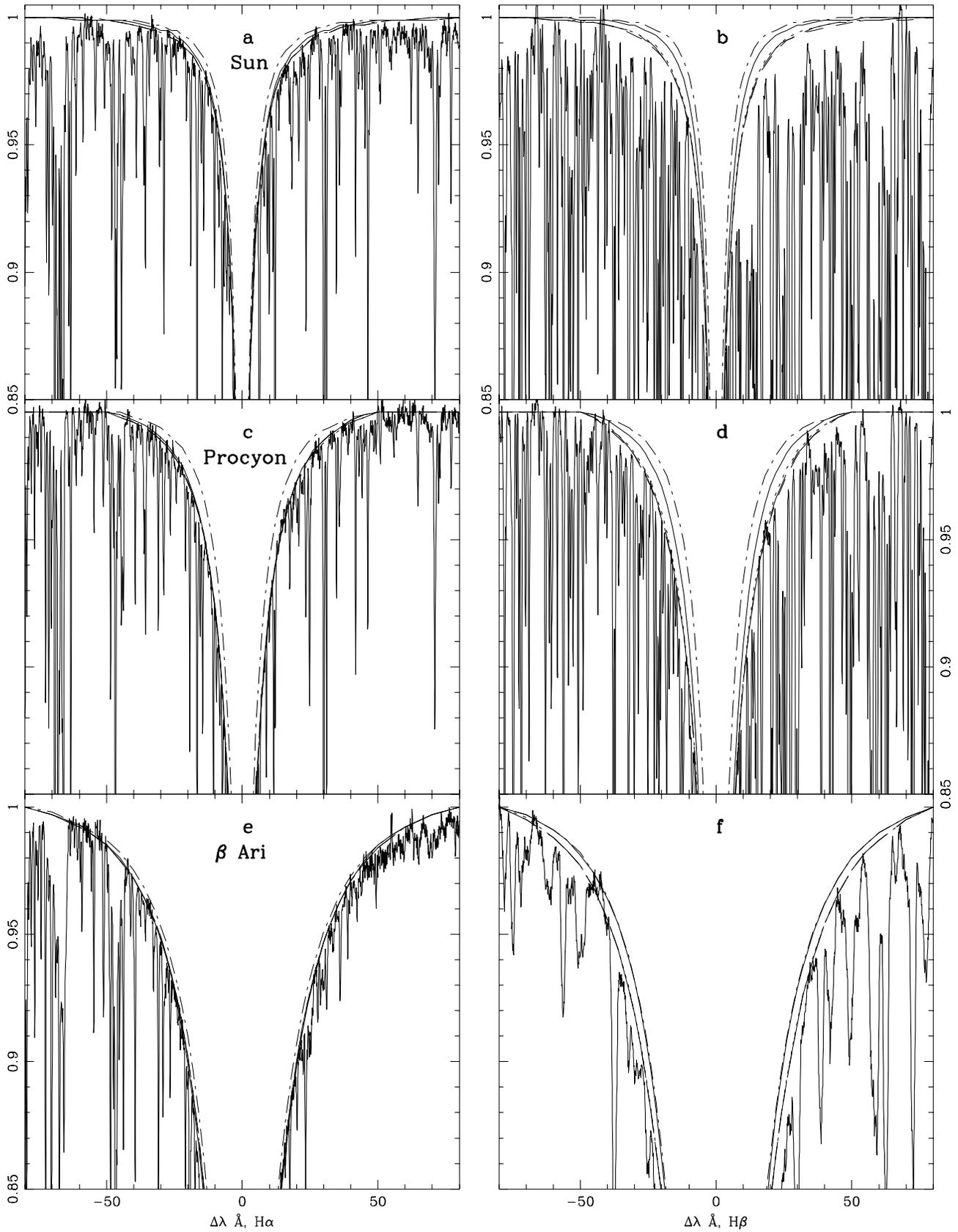}}
  \caption{H$_{\alpha}$ (left panels) and H$_{\beta}$ (right panels) line profiles 
  computed with different convection models represented by the same line styles as
  in Fig.~\ref{Ttau_single}, together with the observed profiles for the same three
  stars as in Fig.~\ref{Ttau_single}.}
  \label{BLP_single}
\end{figure*}

The effect of changing  the physical parameters entering the models on the Balmer
line profiles (BLPs) is very complex. This is illustrated in Fig.~\ref{BLP_single},
where the synthetic profiles for several different convection models are compared 
to the observed ones for the same three stars as in Fig.~\ref{Ttau_single}. 
The spectra shown in Fig.~\ref{BLP_single} were obtained at the Haute-Provence
Observatory, with the spectrograph Aur\`elie attached to the 152cm reflector,
equipped with a CCD receptor. The resolution is about 25000.
The Aur\`elie spectra are observed in the first or second order, 
depending on the wavelength. The wavelength range is 200 \AA,
and the continuum tracing is local, using the most suitable windows.
With a signal to noise ratio of at least 400 we can expect an accuracy for the continuum
location of 0.3 \%, i.e.~0.5 \% for the ratio of line to continuum fluxes in the line wings.
This corresponds to a 30 to 60~K change in effective temperature for F to G stars.

\begin{figure*}
\resizebox{\hsize}{!}{\includegraphics[bb= 85 365 500 720]{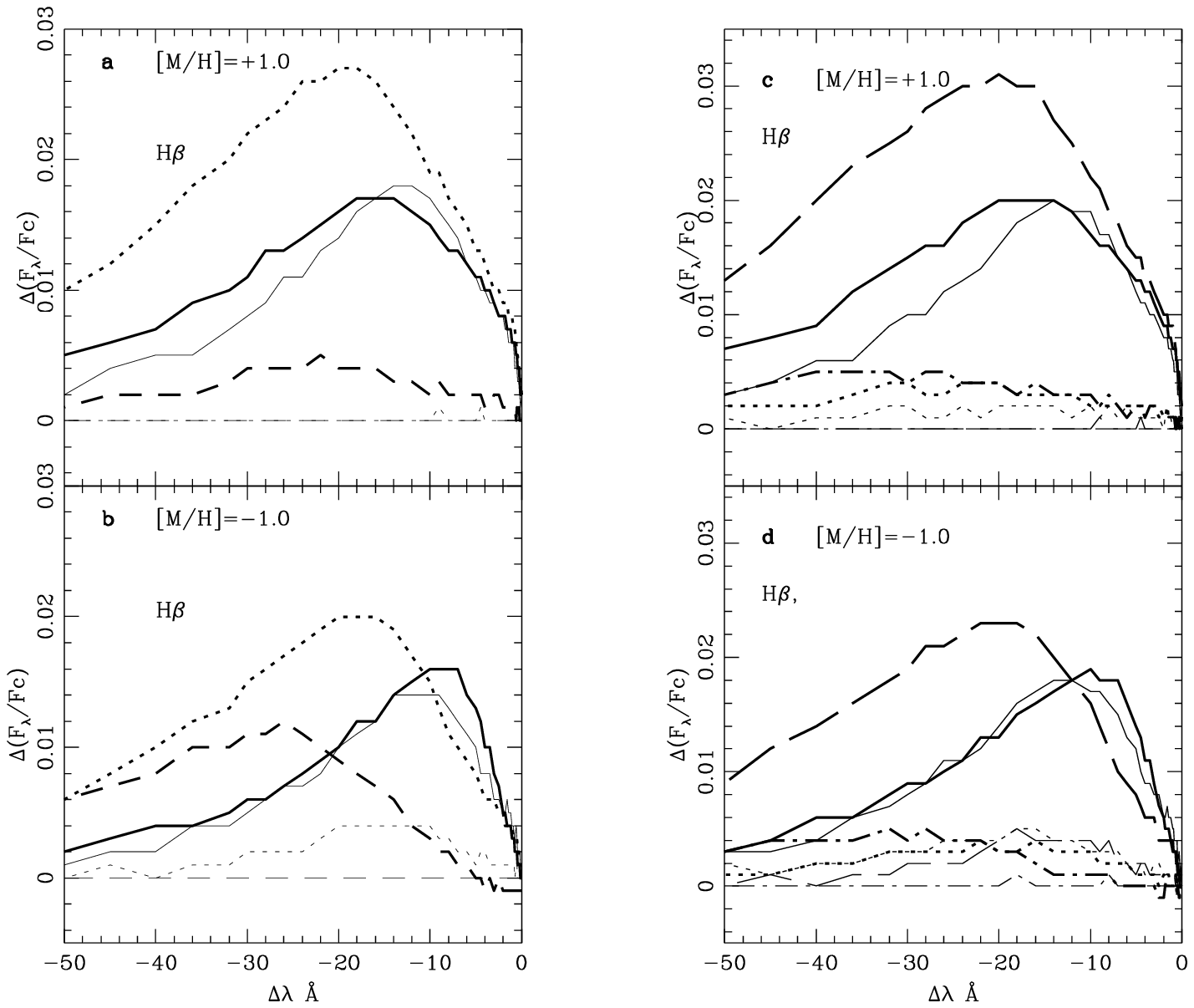}}
  \caption{Differences between normalized fluxes in Balmer line profiles
  obtained from two models differing only by the convection treatment, versus the
  distance from the line center $\Delta \lambda$ in \AA ngstrom units. Fluxes are
  normalized to the local continuum flux $F_{\rm c}$. The two left panels display the
  difference of fluxes computed using MLT models differing by the $\alpha$ 
  parameter: $F(\alpha=1.25)-F(\alpha=0.5)$ for two metallicities. 
  The different line styles correspond
  to different \Teff's: full lines for 6500~K, dotted lines 7500~K and 
  short-dashed ones 8500~K, and thin and thick lines respectively correspond to
  \logg\ = 2.5 and 4.5. In the right panels the models differ
  by the convection formulation, and are identifiable as follows: for
  MLT($\alpha=1.25) -$ CM full lines correspond to \Teff\ = 6500~K, long dashed lines
  to \Teff\ = 7500~K;  for MLT($\alpha=0.50) -$ CM dotted lines correspond to
  \Teff\ = 6500~K, and short-dash-dotted lines to \Teff\ = 7500~K. Thin and thick
  lines have the same meaning.}
  \label{BLP_alpha}
\end{figure*}

In the case of H$_{\alpha}$ the effects are never larger
than 0.5\%. Fig.~\ref{BLP_single} shows 
that the H$_{\alpha}$ profile is insensitive to 
the choice of any of the scale lengths or convection models discussed
in Sect.~\ref{convection}, while in the case of H$_{\beta}$ the
profiles computed with MLT and $\alpha = 1.25$ are too narrow. As an example,
in the case of Procyon this H$_{\beta}$ profile must be computed with \Teff\ 
around 300~K higher to represent the observed profile. 
The insensitivity of H$_{\alpha}$ to any convection treatment is one of the
reasons why it is a very good temperature indicator. However, it is formed close
to the boundary of convectively instable layers and therefore can be modified
by inclusion of overshooting. 

Fig.~\ref{BLP_single} is a convincing illustration that by the use of the CM or CGM
convection treatment, the observed H$_{\alpha}$ and H$_{\beta}$ profiles can be represented
by the same atmosphere model, and this constraint can be achieved in the MLT case
provided a value for $\alpha$ of about 0.5 is adopted. Thus, we want to emphasize
that for the three stars investigated here, {\it less efficient convection within
ATLAS9 type model atmospheres allows the best fit of H$_{\alpha}$ and H$_{\beta}$ using
the same atmosphere model.}

In the MLT case, the consequences of these effects on the BLPs have been
extensively described by \citetalias{cvm} and \citet{fag93}, who demonstrated that
the BLPs are sensitive probes of the atmosphere structure and of effective
temperature for late A, F, and G dwarf stars. Fig.~\ref{BLP_alpha}a,b
illustrates the effect of changing the MLT parameter $\alpha$ on the BLPs, and to
what extent this modification depends on the model parameters. This sensitivity
strongly depends on the selected combinations of the model parameters.
For instance, the largest differences are seen for models with \Teff\ between
7000K and 8000K, \logg\ = 4.5 and high metallicity. In contrast, the  differences
are insignificant at low gravity, high metallicity and high temperature. The shape
of the profiles is also affected, the most for low temperature and low metallicity 
models.

Fig.~\ref{BLP_alpha}c,d shows the difference between two H$_{\beta}$ line profiles,
computed with MLT and CM. The difference between the CGM and CM H$_{\beta}$ profiles was  
not plotted, because it is similar to MLT($\alpha = 0.5$) $-$ CM.
The differences with MLT($\alpha = 1.25$) are the largest and strongly depend on
temperature and gravity, but less on metallicity. These statements mean that
the most efficient convection treatment is MLT with $\alpha = 1.25$, in agreement
with the $T-\tau$ laws shown in Figs.~\ref{Ttau_single} and \ref{Ttau_general}.
From the observer's point of view, Fig.~\ref{BLP_alpha}a-d 
also reveals that Balmer line profiles have to be measured and 
normalized to an accuracy of at least 0.5\% to draw a clear 
distinction between convection models with different efficiency.
Insufficiently determined profiles may thus easily introduce erroneous
trends or large scatter when analyzing their dependence on a particular
convection treatment.

\begin{figure}
\resizebox{\hsize}{!}{\includegraphics[bb= 80 205 310 715]{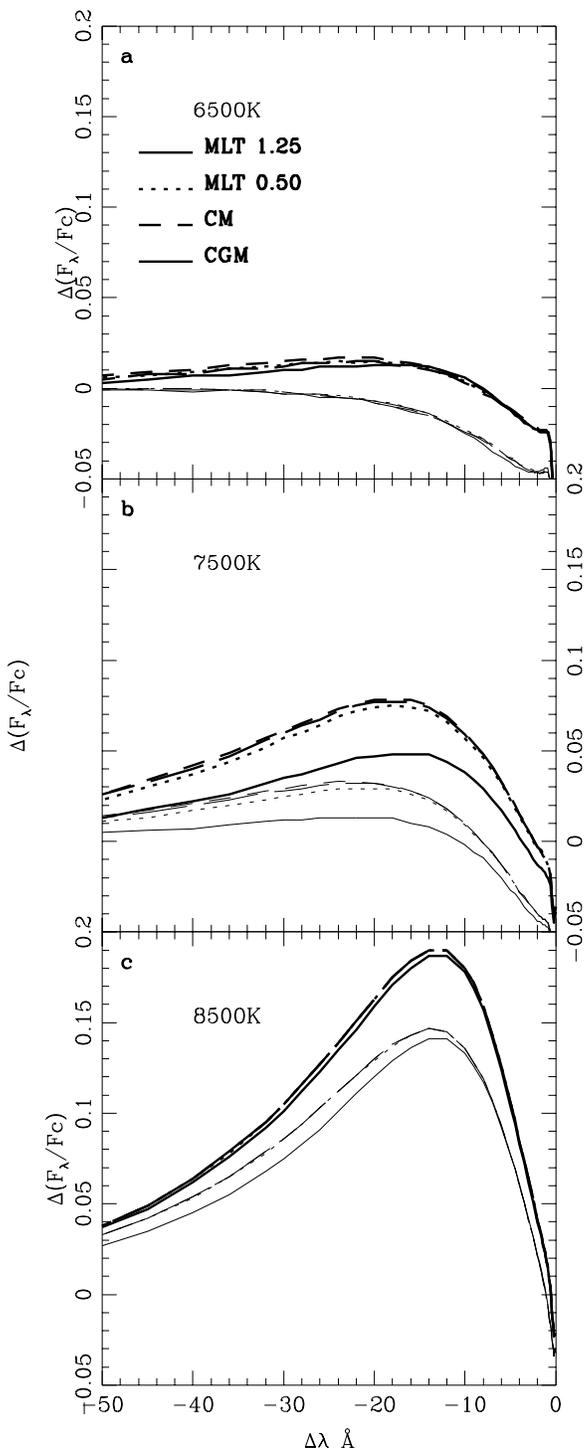}}
  \caption{Differences between H$_{\beta}$ line profiles computed with \logg\ = 2.5
  and 4.5. Different convection models are represented by different line styles, for
  metallicities of $-$1 and +1 by thin and thick lines respectively, and for
  three effective temperatures.}
  \label{BLP_logg}
\end{figure}

We stress here that the sensitivity to MLT's parameter $\alpha$ 
strongly depends on gravity. For instance, for \Teff\ $\le$~7500~K all
convection models with low gravity yield inefficient convection. This is due to the
fact that low gravity implies lower densities, and the convective efficiency is
related to the density as explained in subsection 5.1 above.
Moreover, we suggest to consider the {\em commonly assumed insensitivity of
BLPs to gravity change for \Teff\ below 8000~K with real caution}. Indeed,
Fig.~\ref{BLP_logg} shows clearly that the sensitivity of BLPs to gravity changes
depends more than usually expected on metallicity, \Teff, and finally on all
parameters playing a role in the efficiency of the convective transport. It depends
also on the gravity itself, the second derivative is not zero. This effect is most
important for the highest metallicity and largest \Teff. The main
reason is that an increase of metallicity leads to a lowering of the density on
the one hand while a higher effective temperature favors radiative transfer on the
other. 

\subsubsection{Fluxes and colors}
\label{flux_col}

\begin{figure*}
\resizebox{\hsize}{!}{\includegraphics{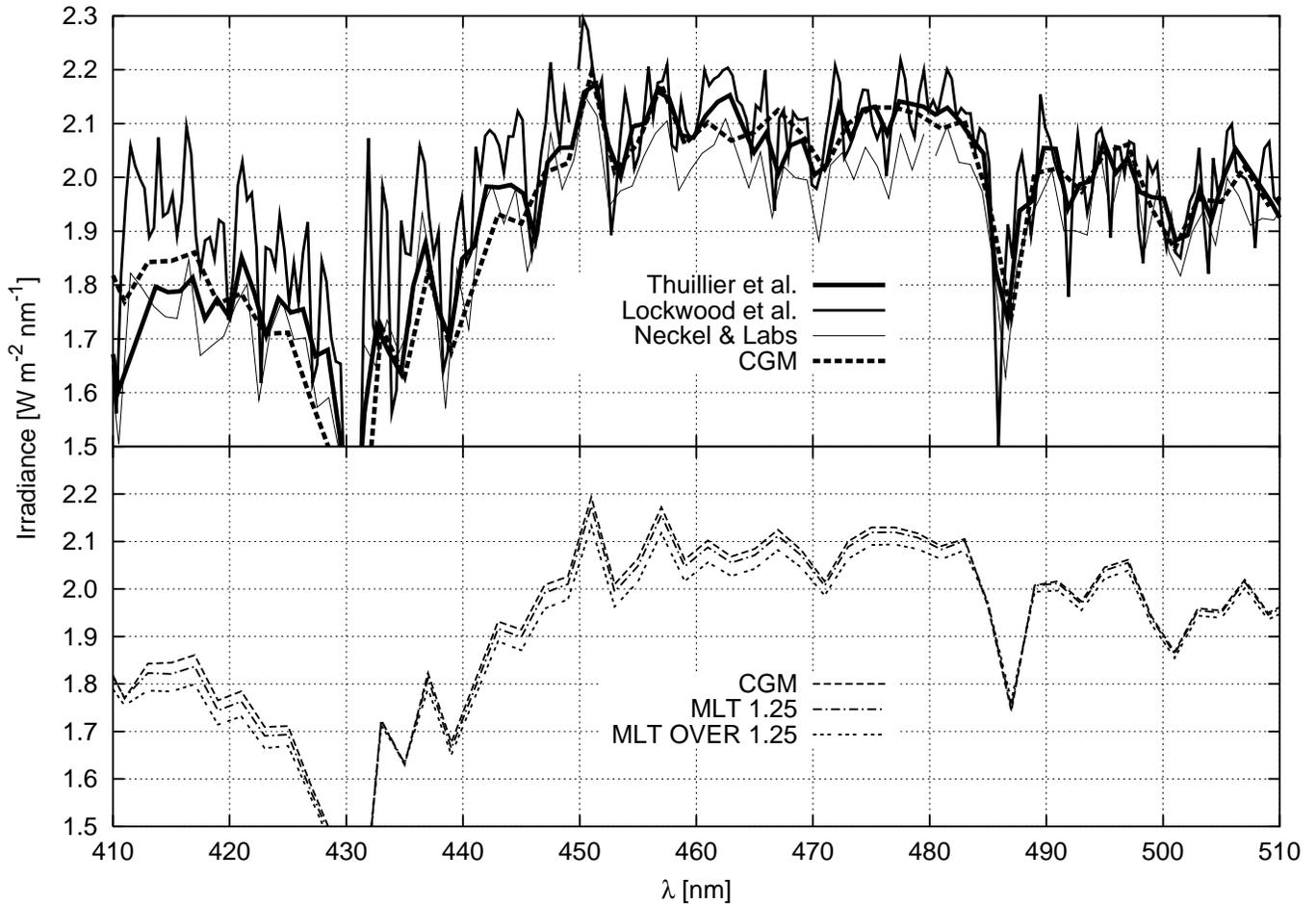}}
  \caption{{\em Top:} Observations of the solar irradiance 
  from \citet{Thui:98}, \citet{Lock:92} and \citet{Neck:84} (solid lines)
  and solar irradiance calculated with the CGM model (dashed line).
  {\em Bottom:} Solar irradiance calculated with three different convection models.}
  \label{irradiance}
\end{figure*}

We have computed fluxes for solar models with different convection treatments
as follows: CGM, CM, MLT ($\alpha$=0.5), MLT ($\alpha$=1.25), MLT ($\alpha$=1.25 
with overshooting). The same parameters have been chosen for all models:
\Teff\ = 5777~K, \logg\ = 4.4377, \Vmicro\ = 1.5~\kms, element abundances from
\citet{Ande:89}, except for Fe, for which the current value of
$\log(N_{\rm Fe}/N_{\rm tot}) = -4.51$ was used \citep{kuruczII}.
To compare the calculated solar fluxes to observations, solar irradiance data 
have been taken from \citet{Neck:84}, and in addition from two more
recent sources: \citet[ Lowell Observatory, 1985]{Lock:92} and 
\citet[ SOLSPEC spectrometer on ATLAS I mission, 1992]{Thui:98}.
The irradiances in the region of maximum emitted radiation, i.e.\ 
410 to 510~nm, are displayed in Fig.~\ref{irradiance}. 

As can be seen, the three
observational data sets are different from each other by up to 15~\% (upper panel),
although \citet{Neck:84} estimate 0.5~\% as an upper limit for the local systematic error
of their measurements. But they point out that intrinsic intensity variations
depending on solar activity can occur when comparing measurements made at different
times. These amount to 2~\% in certain spectral regions (e.g.\ the CaII K line) in their data,
which are derived from observations made over a 20~yr period \citep[see also][]{Livi:91}.
For comparison, \citet{Lock:92} give an upper limit for the errors of their measurements
of 2~\% (their observations were made at a phase of low solar activity), and \citet{Thui:98}
quote a mean uncertainty of 2-3~\% (data obtained at high solar activity)
\footnote{Information on the solar activity level has been obtained from the
National Solar Observatory Digital Library (http://www.nso.noao.edu/diglib/ftp.html).}.
Detailed discussions of the error sources and comparisons with previous observations
are given in each of the three references.

However, the mean of the maximum relative difference between the irradiances from the three sources
is 5~\% in the region of 450 to 480~nm, 
which is much larger than the differences between the fluxes calculated with
different convection models (2~\%, cf.\ lower panel of Fig.~\ref{irradiance}). 
The CM and MLT ($\alpha$=0.5) fluxes are
almost identical to the CGM flux. Therefore, the solar irradiance measurements cannot be
used to decide between the various models. A similar conclusion would result if 
measurements of solar central intensity would be used, because the error estimates
by \citet{Neck:84} for these measurements (the other two sources did not include this kind 
of measurements) are equal to that for the irradiance spectra.
Thus, we regard tests of central intensity calculations against observations
\citep[e.g.][]{cast97} as having limited significance until accuracy and
absolute calibration of these data will have been established with the necessary reliability.

\begin{figure}
\resizebox{\hsize}{!}{\includegraphics[bb=80 411 540 705,clip]{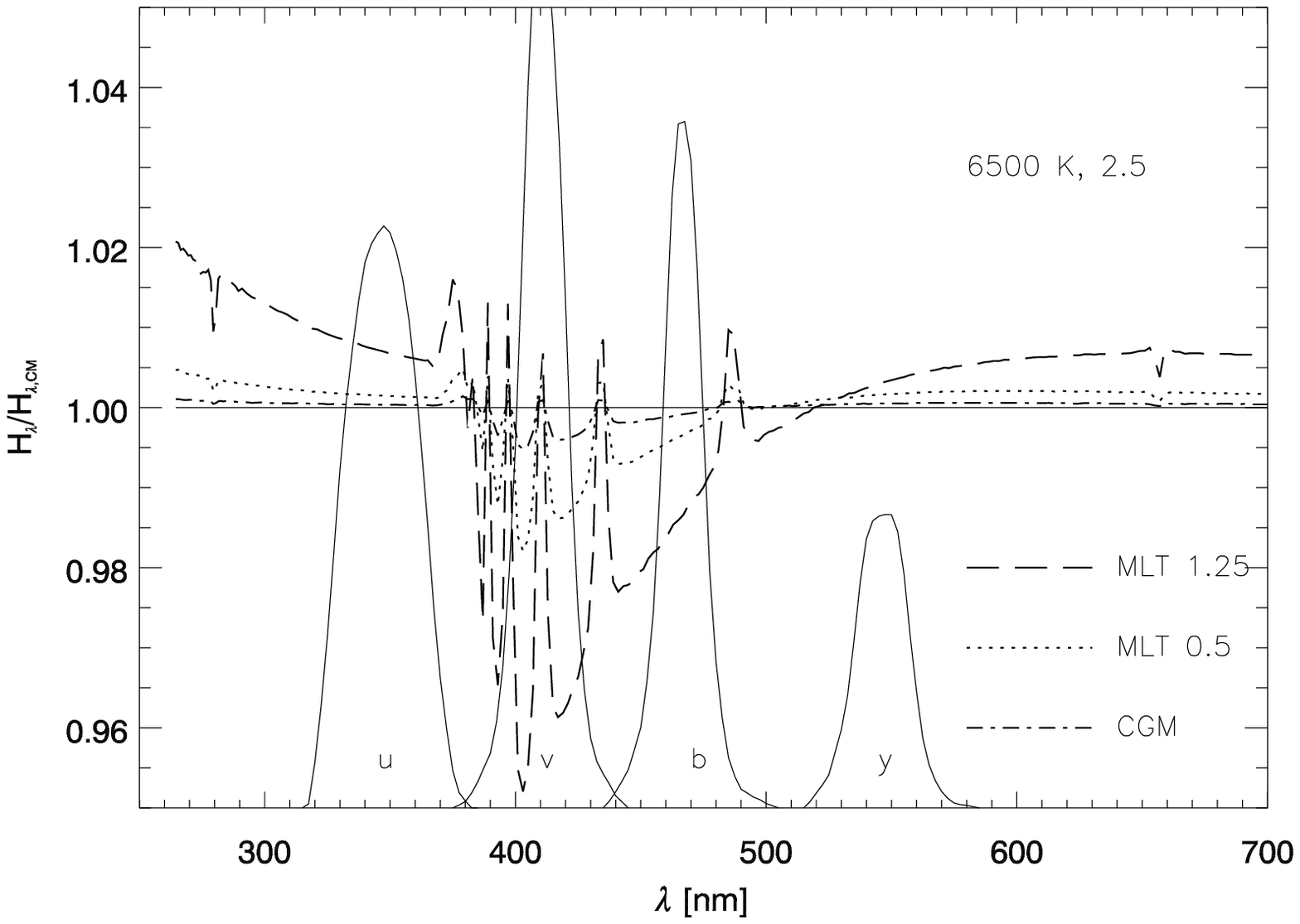}}
\resizebox{\hsize}{!}{\includegraphics[bb=80 411 540 694,clip]{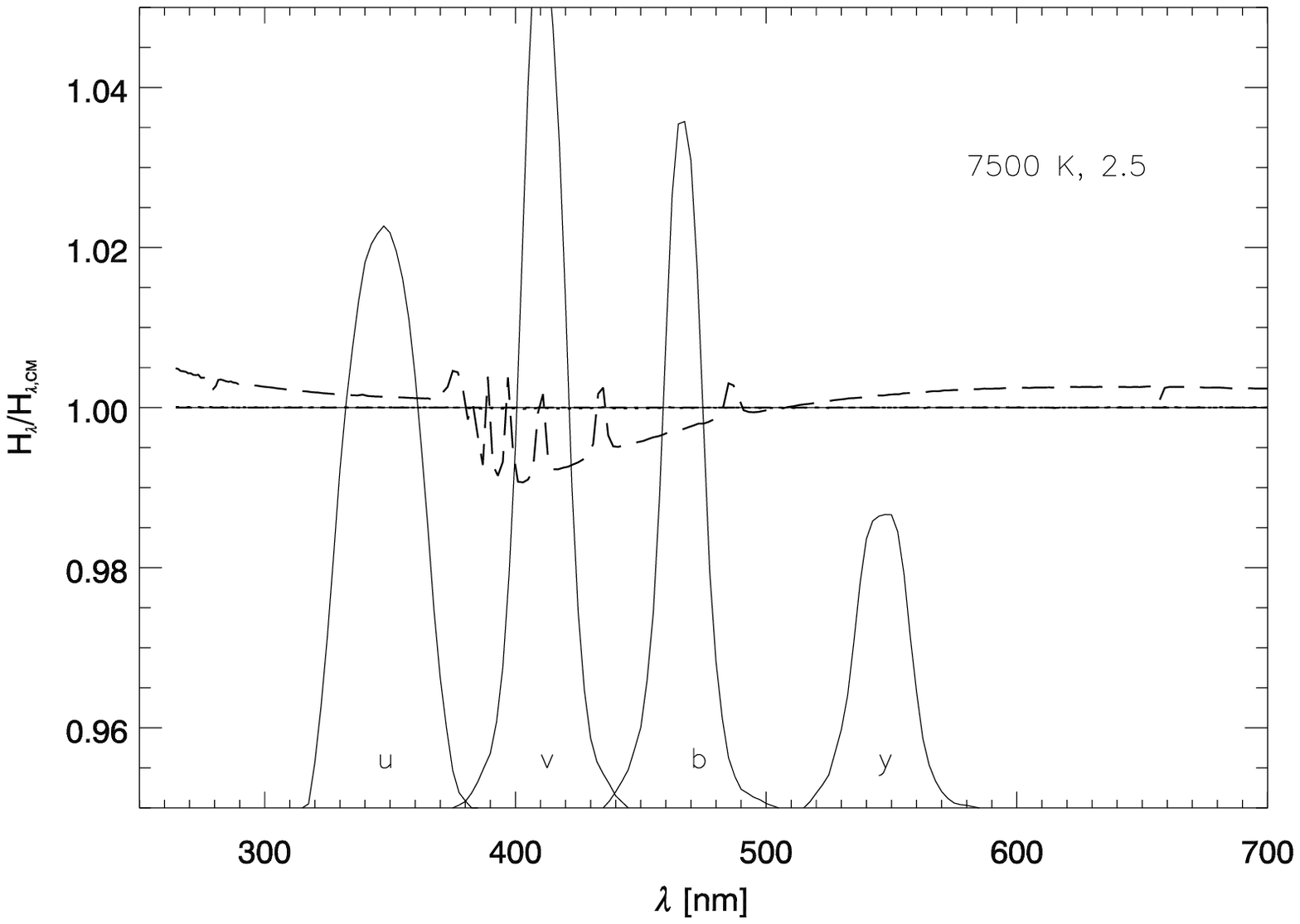}}
\resizebox{\hsize}{!}{\includegraphics[bb=80 370 540 694]{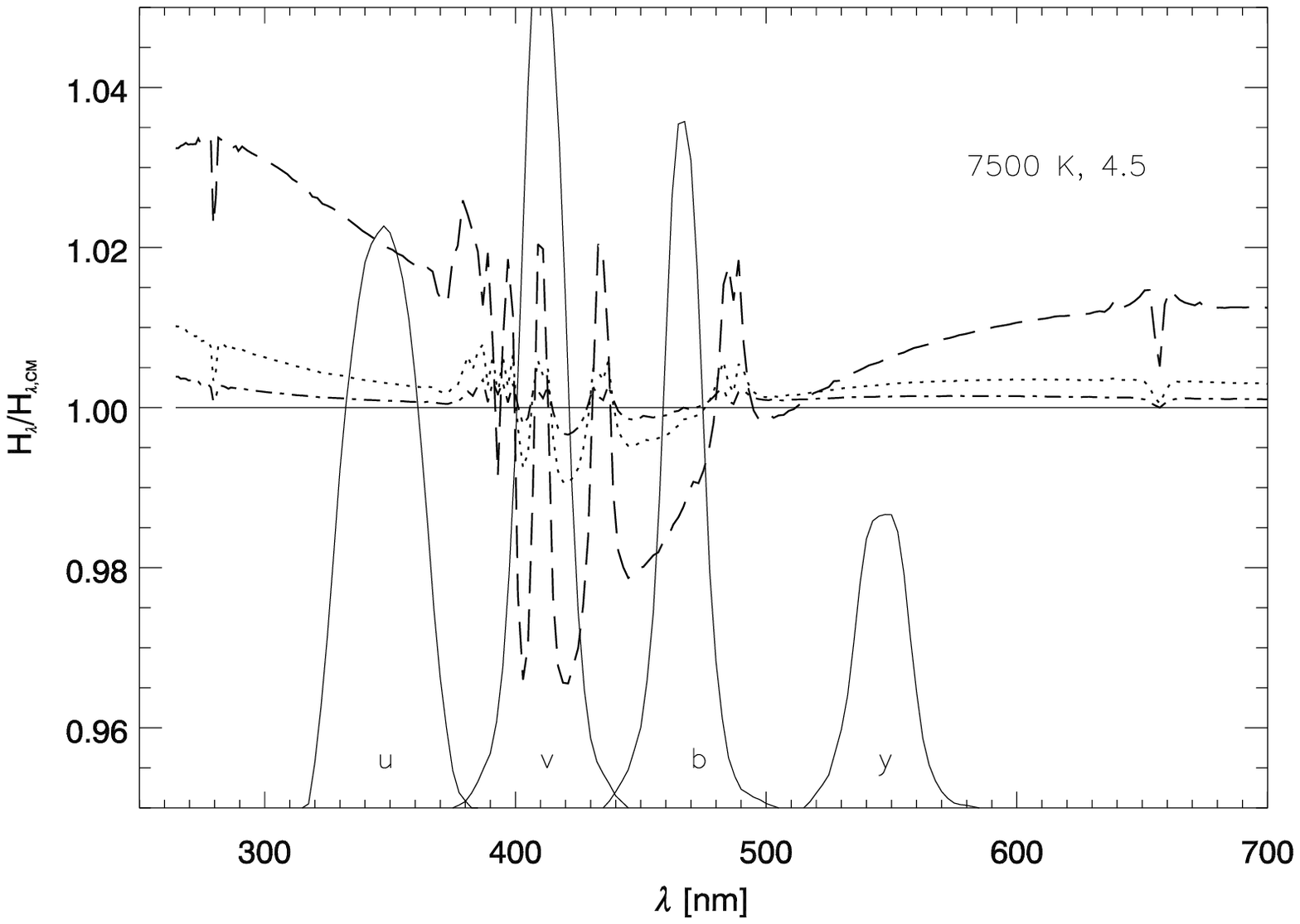}}
  \caption{Ratios of MLT and CGM to CM fluxes for three different combinations
  of \Teff\ and \logg. \logZ=$-$1 for all cases. Relative transmissivity 
     profiles of the Str\"omgren $uvby$ filters are indicated (multiplied by a
     standard 1P21 photomultiplier response function).}
  \label{flux_fig}
\end{figure}

The general dependence of the calculated flux on the convection model
can be seen in Fig.~\ref{flux_fig}, where the ratios of MLT and CGM 
to CM fluxes are displayed for two different values of \Teff\ and \logg\
and the extreme case of \logZ = $-$1. For all cases, the CGM flux
is closest to the CM flux, followed by MLT ($\alpha$=0.5) and with a larger
discrepancy by MLT ($\alpha$=1.25). The differences between the
models are very small for the highest \Teff\ and lowest \logg\ values.
Otherwise they depend strongly on the wavelength
range, and no general trend is visible. This is illustrated by Table~\ref{flux_tab},
which lists the (\Teff, \logg) combinations for three metallicities in
order of increasing flux differences from top to bottom. Three different wavelength
ranges have been regarded: blue, UV and red. It can be seen that
in the latter two, the convective efficiency effect is inversed compared to the 
one in the blue part.
Thus, one can only guess that the calculated emitted flux depends in a complex way
on the combination of \Teff, \logg, and the convection model.

\begin{table}
\caption{Model parameters ordered according to increasing flux differences
arising when using any two different convection models.
The quantities listed in the columns labeled ``blue'', ``UV'' and ''red''
are the maxima of $|H_{\lambda,{\rm MLT 1.25}}/H_{\lambda,{\rm CM}}-1|$
for the wavelength ranges 360--520~nm, 250--360~nm and 520--700~nm, 
respectively (cf.\ Fig.~\ref{flux_fig}).}
\label{flux_tab}
\begin{tabular}{llllllll}
\hline\hline
blue & \Teff & \logg & & UV & red & \Teff & \logg \\
\hline
\multicolumn{8}{c}{\logZ = +1} \\
\hline
 0.000 & 7500 & 2.5 & & 0.000 & 0.000 & 7500 & 2.5 \\
 0.024 & 6500 & 4.5 & & 0.011 & 0.005 & 6500 & 4.5 \\
 0.038 & 6500 & 2.5 & & 0.012 & 0.007 & 6500 & 2.5 \\
 0.046 & 7500 & 4.5 & & 0.022 & 0.011 & 7500 & 4.5 \\
\hline
\multicolumn{8}{c}{\logZ = 0} \\
\hline
 0.004 & 7500 & 2.5 & & 0.003 & 0.001 & 7500 & 2.5 \\
 0.030 & 6500 & 4.5 & & 0.014 & 0.007 & 6500 & 4.5 \\
 0.043 & 7500 & 4.5 & & 0.022 & 0.008 & 6500 & 2.5 \\
 0.048 & 6500 & 2.5 & & 0.027 & 0.013 & 7500 & 4.5 \\
\hline
\multicolumn{8}{c}{\logZ = $-$1} \\
\hline
 0.009 & 7500 & 2.5 & & 0.006 & 0.003 & 7500 & 2.5 \\
 0.026 & 6500 & 4.5 & & 0.023 & 0.007 & 6500 & 2.5 \\
 0.034 & 7500 & 4.5 & & 0.030 & 0.011 & 6500 & 4.5 \\
 0.048 & 6500 & 2.5 & & 0.034 & 0.015 & 7500 & 4.5 \\
\hline\hline
\end{tabular}
\end{table}

We use our grids of computed fluxes to derive colors and color indices in the
$uvby$ photometric system. The role of convection on this photometric system has
already been studied by SK (see Sect.~\ref{previous}), also using the ATLAS9 code 
in the MLT and CM cases. Here, we extend this study to the CGM
case, and investigate  how the variations of color indices due to temperature
and gravity variations are affected by the convection formulation. 

We find that:
\begin{itemize}
\item there are no measurable differences between colors or indices computed with CM and 
      CGM models and both are very close to those computed with 
      MLT($\alpha$=0.50) models.
\item differences become much more important if colors or indices computed 
      with MLT($\alpha$=1.25) models are compared to those computed with 
      CM, CGM, or MLT($\alpha$=0.50) models.
\end{itemize}

Thus, we can extend the results of SK (see Sect.~\ref{previous}) to the CGM model,
and conclude that color indices computed with CGM models are generally in better agreement with
observations than those computed with MLT($\alpha$=1.25) models.

\begin{figure}
\resizebox{\hsize}{!}{\includegraphics[bb= 50 185 205 325]{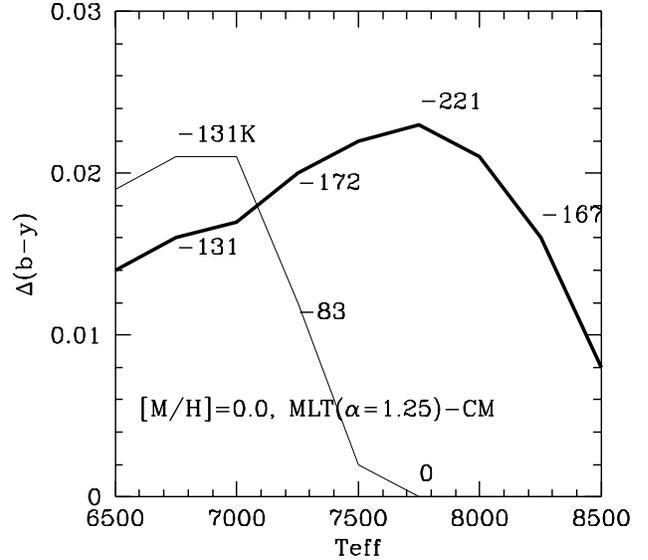}}
  \caption{Differences between $(b-y)$ indices computed using 
  models differing only by the convection treatment. The thin line is for 
  $\log g=2.5$, the thick line for $\log g=4.5$. For clarity, we indicate 
  the temperature differences corresponding to the $(b-y)$ differences 
  along the curves only for a few models.}
  \label{col_fig}
\end{figure}

The $(b-y)$ index is the most sensitive one with respect to temperature
changes, and this sensitivity is also strongly influenced by the 
convection model considered. We have investigated the variation of
$(b-y)$ indices computed using models differing only by the
convection treatment. Fig.~\ref{col_fig} shows that the 
sensitivity of $(b-y)$ to convection change is \Teff\ and gravity dependent,
and the temperature changes associated 
with the ones of $(b-y)$ are written along the curves.
The results are very similar, and in the same order of magnitude, for 
metallicities ten times or one tenth of the solar one. The same conclusion
is reached when the CM model is replaced by the CGM or by MLT($\alpha$=0.50)
formulation.

From this result we can establish that the ``error" (or ``change") on 
temperature variation estimations can be important, i.e.\ as large as 200~K, when
using MLT($\alpha$=1.25) instead of CM convection treatment (or CGM or
MLT($\alpha$=0.50)). It can reach up to 400~K, if the overshooting option of ATLAS9
is not removed.

\section{Conclusions}

One of the main conclusions to be raised from this study is that as long as
one considers inefficient convection, whatever is the choice of the formulation, 
either MLT with low $\alpha$, or FST, the interpretation of spectroscopic or
photometric observations is equivalent: 
observed BLPs and Str\"omgren color indices of dwarf and subgiant stars
between A5 and G5 spectral types, and in a large range of metallicity
are best represented by the use of less efficient convection transport, i.e.\ MLT
with $\alpha = 0.5$, or with FST formulation. This confirms results already obtained
by \citet{fag93}, \citetalias{cvm} and \citet{Veer:98} for the Sun, Procyon, and other cool metal-poor stars using MLT models.

\citet{Gard99} reported a few opposite cases (see Sect.~\ref{previous}), 
but for parts of their sample of stars fundamentally known \logg\
values were not available. An analysis of a larger sample of stars 
in binary systems with revised fundamental parameters for {\em both}
\logg\ and \Teff\ \citep{Smal:02} did not confirm the
discrepancies previously found. Furthermore, for the case of F stars
\citet{Smal:02} noticed a larger systematic difference between
fundamental effective temperatures and those obtained from H$_{\beta}$
lines for MLT($\alpha$=1.25) than for less efficient convection models,
although this discrepancy remains within the overall uncertainties.

Nevertheless, we have to emphasize that in models with deep convection zones
(e.g.\ for Sun, Procyon) MLT($\alpha = 0.5$) and FST treatments
have comparable effects on calculated fluxes, but not on atmosphere
structure. They produce different temperature gradients in the deep layers,
as can be seen in Figs.~\ref{Ttau_single} and \ref{Ttau_general},
but those cannot be distinguished by the computed BLPs. 
In other words, the BLPs allow to discriminate among different
values for the MLT parameter $\alpha$, but not among MLT($\alpha = 0.5$), CM, 
and CGM models. In any case, the sensitivity of BLPs to convection parameters
depends significantly on the other physical parameters. This holds especially
for the sensitivity to gravity change, which can be more important than usually
expected.

In case of weakly efficient convection, 
fluxes and colors depend only weakly on the selected convection treatment. 
On the other hand, when the convection is highly efficient, then fluxes and colors
become strongly dependent on the convection modelling, as the differences
among the models show up more clearly within the photosphere.
Thus, significant uncertainties on stellar global parameters 
arise from the convective treatment in model atmospheres. Ignoring these
uncertainties can lead to systematic differences affecting subsequent interpretations.

The calculations of color and limb darkening partial derivatives 
are significantly improved when using the present model atmosphere
grids which are finer spaced in \Teff\ and \logg\ and have
a higher resolution in the temperature
distribution with depth \citep{Bar2002}. Smoothness of these derivatives is
of crucial importance in the mode discrimination problem for non-radially
pulsating stars, which is basically
due to the dependence of the color amplitude ratios on these derivatives.
Details of the required precision of these partial derivatives in order to be
useful for mode identification will be given in \citet{Gar2002}. 
There we will show that the next space asteroseismology missions -- COROT,
MONS/R\o mer and Eddington -- will supply light curves with high enough precision
to permit a direct comparison up to the second order to partial derivatives with
respect to temperature and gravity as calculated with the present model atmospheres.

The improved resolution of the new model grids also avoids unphysical
oscillations in evolutionary track calculations when using ATLAS9
model atmospheres as boundary conditions (see Sect.~\ref{resolution}). Moreover, 
the possibility to choose among different convection models allows a self-consistent
match between model atmospheres and model envelopes \citep{Mont2001}.
However, we must stress here that the different relations 
$T$ and $F_{\rm conv}/F_{\rm tot}$ vs.\ depth represent stars which are different 
in their radii and luminosities. 
The broad effects of the convective treatment can only be assessed 
by studying a complete stellar model, i.e.\ a model with an atmosphere and 
an internal structure which are consistently built with the same convection formulation.
We will address this topic in follow up work \citep{Kup2002}.

\begin{acknowledgements}
The authors would like to thank Gerard Thuillier for providing the
SOLSPEC solar irradiance data. 
Many thanks go to Robert Kurucz for allowing us to use his model atmosphere code
and opacity data.
We would like to thank the referee, F. Castelli, for helpful comments and the rapid 
evaluation of the manuscript.
This research was carried out within the working group {\em
Asteroseismology--AMS}, supported by the Fonds zur F\"orderung der 
wissenschaftlichen Forschung (project {\sl P13936-TEC}).
\end{acknowledgements}

\section*{Appendix: Description of SMGT}

The program can be run in either of
two modes depending on the temperature structure used for initialization:
\begin{itemize}
\item In the {\em static} mode, an existing model file or a gray atmosphere is used.
\item In the {\em dynamic} mode, an existing model file or a weighted average of 
all existing models within one grid step of each parameter is used. The weights 
are calculated in the following way:
For each atmospheric parameter $p$, the quantity
\begin{equation}
\exp\left(-\frac{|p^i-p^m|}{p_{\rm max}-p_{\rm min}}\right)
\end{equation}
is computed,
where $i$ denotes the initialization model, $m$ the model to be computed,
and $p_{\rm max/min}$ the maximum and minimum of the parameter as given in
the grid definition.
The results for all parameters (at most four) are multiplied to yield the weight.
\end{itemize}

The state of convergence of a particular model is measured by calculating
the RMS and the maximum of the flux errors ($\Delta F_{l}$), 
the flux derivative errors ($\Delta F'_{l}$) and the temperature correction 
($\Delta T_{l}$) in all layers $l$. A model is considered as fully converged
if these values satisfy certain criteria, which are given in 
Table~\ref{conver_crit} (labeled ``primary''). Models for which
these criteria cannot be achieved after a reasonable number of
iterations are also stored, if they satisfy the criteria labeled ``secondary''
in Table~\ref{conver_crit}, but the corresponding files are marked with ``$\sim$''.

\begin{table}
\caption{Convergence criteria used for the Vienna grid computations.}
\label{conver_crit}
\begin{tabular}{lrrc}
\hline\hline
                            &        RMS &  Maximum  & \\
\hline
primary & & & \\
\hline
 $\Delta F_{l}$               & $\le$ 1\%  & $\le$ 5\% & \\
\hline
 $\Delta F'_{l}$              & $\le$ 2\%  & $\le$ 10\% & \\
                              &            &  or $\le$ 10\% & ($N/10\le l \le N$) \\
                              &            &  and $\Delta T_{l}$ $\le$ 1K & ($0\le l \le N/10$) \\
\hline
secondary & & & \\
\hline
 $\Delta F_{l},\Delta F'_{l}$ & $\le$ 10\% & $\le$ 100\% & \\
\hline\hline
\end{tabular}
\end{table}

In order to achieve the convergence criteria 
without wasting time when no further improvements can be expected from 
further iterations, the required number of iterations is calculated and checked
dynamically, after an initial sequence of 12 iterations. From a sequence of
$n$ iterations the ATLAS9 output is processed every $n$/4 
iteration (but at least every 15$^{th}$ and at most every 3$^{rd}$
iteration) and the speed of convergence is characterized in the following way.
The ratios between the rms errors of two subsequent iterations
($r_F^i, r_{F'}^i$) are calculated. If they are found smaller than a
threshold value (0.95), damping exponents are computed
iteratively for the flux errors:
\begin{equation}
\gamma_F^i = \left[\gamma_F^{i-1} n^{i-1} - a \ln(1-r_F^i)\right]/n^i,
\end{equation}
where $i$ goes from 2 to the number of processed outputs, 
$\gamma_F^1$ is set to zero or the value determined from the previous iteration
sequence, $a = 1$ for $i = 2$ and $a = (n^i-n^{i-1})/n^i$ for $i > 2$,
and an analogous formula is used for the flux derivative errors.
The total number of iterations is then given by
\begin{eqnarray}
n_{\rm tot} & = & {\rm int}\left(\frac{\ln(r_{\rm crit})}{\ln(1 - \exp(-\gamma_F n))}\right) \nonumber\\
{\rm or} \quad n_{\rm tot} & = & {\rm int}\left(\frac{\ln(r_{\rm crit})}{\ln(r_F)}\right)
\end{eqnarray}
with
\begin{equation}
r_{\rm crit} = \frac{{\rm RMS_{prim}}(\Delta F_{l})}{{\rm RMS}(\Delta F_{l})}
\end{equation}
where the largest of the values resulting from $\Delta F_{l}$ or $\Delta F'_{l}$ is taken.
Apart from the first iteration sequence, 
the actual number of further iterations is a fraction depending on the ratio 
between the current predicted number of total iterations and the previous 
prediction.
The computations are terminated if the errors increase (after a few further
trials) or if the total number of iterations would be too large
(we use a limit of 480).

\bibliography{models}
\bibliographystyle{aa}
              
\end{document}